\documentclass[preprintnumbers,amsmath,amssymb,superscriptaddress,pra,notitlepage]{revtex4-1}

\include{amsmath}
\usepackage{graphicx}
\usepackage{subfigure}
\usepackage{amsmath}
\usepackage{amsfonts}
\usepackage{color}
\usepackage[latin1]{inputenc}
\usepackage{times}

\begin{document}

\title[]{Theory of bifurcation amplifiers utilizing the nonlinear dynamical response\\ of an optically damped mechanical oscillator}
\author{Kjetil B{\o}rkje}
\affiliation{Department of Science and Industry Systems, University of South-Eastern Norway, PO Box 235, Kongsberg, Norway}
\date{\today}

\begin{abstract}
We consider a standard optomechanical system where a mechanical oscillator is coupled to a cavity mode through the radiation pressure interaction. The oscillator is coherently driven at its resonance frequency, whereas the cavity mode is driven below its resonance, providing optical damping of the mechanical oscillations. We study the nonlinear coherent response of the mechanical oscillator in this setup. For large mechanical amplitudes, we find that the system can display dynamical multistability if the optomechanical cooperativity exceeds a critical value. This analysis relates standard optomechanical damping to the dynamical attractors known from the theory of optomechanical self-sustained oscillations. We also investigate the effect of thermal and quantum noise and estimate the noise-induced switching rate between the stable states of the system. We then consider applications of this system and primarily focus on how it can be used as bifurcation amplifiers for the detection of small mechanical or optical signals. Finally, we show that in a related but more complicated setup featuring resonant optomechanical interactions, the same effects can be realized with a relaxed requirement on the size of the mechanical oscillations.
\end{abstract}

\maketitle

\section{Introduction}
Research on optomechanical systems is of relevance to gravitational wave detection \cite{McClelland2011LasPhotRev}, signal processing \cite{Metcalfe2014ApplPhysRev}, quantum information processing \cite{Reed2017NatPhys}, and the fundamentals of quantum mechanics \cite{Arndt2014NatPhys}. In many such systems, an optical cavity mode and a mechanical oscillator are coupled through the nonlinear radiation pressure interaction. This interaction is generally weak at the single-photon level, with the consequence that, with some exceptions, the equations of motion are effectively rendered linear. Optomechanics in the linear regime has nevertheless enabled remarkable achievements in recent years. Some prominent examples are cooling to the motional quantum ground state \cite{Teufel2011Nature,Chan2011Nature}, creation and detection of quantum entanglement between two remote mechanical oscillators \cite{Riedinger2018Nature,Ockeloen-Korppi2018Nature}, and the realization of nonreciprocal photonic devices \cite{Verhagen2017NatPhys}.

It is well-known that for a particular choice of optical driving, cavity optomechanical systems can display classical nonlinear dynamics even in the regime of small single-photon coupling rate. This can occur when the optical drive is blue-detuned, meaning that its frequency exceeds the cavity resonance frequency. This favors down-conversion of photons through the optomechanical interaction, which tends to amplify mechanical fluctuations \cite{Braginsky2001PhysLettA,Carmon2005PRL,Kippenberg2005PRL}. This amplification is unbounded in a linearized theory, but the analysis of the full nonlinear dynamics predicts large self-sustained oscillations that settle into one of several dynamical attractors \cite{Marquardt2006PRL}. The fact that several different stable oscillation amplitudes exist for the same set of system parameters is referred to as {\it dynamical} multistability, and it means that the system can display hysteresis. The existence of a dynamical attractor diagram has been confirmed in experiments \cite{Krause2015PRL,Buters2015PRA}. It should be noted that there are also stable attractors for red-detuned optical drives, i.e., for drive frequencies below cavity resonance. However, for such detunings, mechanical noise is damped rather than amplified, which means that deliberate driving of the mechanical oscillator is needed in order to reach the attractors where large coherent motion is self-sustained.

Systems that have two or more stable states can for example be useful for signal amplification and for memory storage. In electronics, circuits with this property are commonly referred to as latch circuits, since a signal can cause it to switch to or {\it latch} onto another stable state. Bistability can also arise in optical cavity fields coupled to atoms \cite{Gibbs1976PRL} or in microwave circuits containing Josephson junctions \cite{Siddiqi2004PRL,Siddiqi2005PRL}. The bistable dynamical response of nonlinear nanomechanical oscillators \cite{Dykman1980SovPhysJETP,Aldridge2005PRL,Karabalin2011PRL,Dong2018NatComm} can for example be useful for sensing of external forces. Experiments on optomechanical systems have explored {\it static} bistability where the mechanical system can oscillate around one of two stable equilibrium positions \cite{Dorsel1983PRL,Bagheri2011NatNano,Xu2017NatComm} with potential applications for mechanical memory storage. It has also been suggested that the optomechanical dynamical multistability mentioned above can be useful for sensing, since a small static displacement can cause transitions between two widely different stable oscillation amplitudes in a latching measurement scheme \cite{Marquardt2006PRL}.

In this article, we study the nonlinear response of a mechanical oscillator that is coupled to an optical cavity through the standard radiation pressure interaction. We consider a red-detuned optical coherent beam addressing the cavity, which according to linearized optomechanical theory provides additional damping of the mechanical oscillator. We also assume that the mechanical oscillator is coherently driven at its resonance frequency, which in some cases can be implemented by mechanical actuation, e.g., by piezoelectric elements. However, it may in many cases be more feasible to implement the mechanical drive optically and we show that this is indeed possible. For small mechanical oscillation amplitudes, the mechanical response to the drive is linear and the oscillator's damping rate is indeed enhanced due to the red-detuned optical beam. However, for strong drives and thus large mechanical oscillation amplitudes, the optical damping becomes inefficient. The reason is that the coherent mechanical oscillations cause the cavity resonance frequency to vary, and for large amplitudes these frequency variations can become comparable to the laser detuning itself. By considering all possible oscillation amplitudes, we show that the mechanical response to the drive is highly nonlinear and that, for sufficiently strong optical powers, the optomechanical system can display dynamical multistability. This behaviour is of course related to the dynamical attractor diagram \cite{Marquardt2006PRL} discussed above. Here, we consider the details of addressing this attractor diagram with red-detuned optical drives and thereby connect two well-known optomechanical effects - optical damping and dynamical multistability. 

The nonlinear phenomenon we analyze has similarities with one recently studied experimentally in an electromechanical system \cite{Dong2018NatComm} in the sense that the frictional force experienced by the mechanical oscillator becomes negative only at a sufficiently large oscillation amplitude. In contrast to Ref.~\cite{Dong2018NatComm}, which dealt with an interaction energy depending on oscillator position squared, we consider the standard and ubiquitous radiation pressure interaction which is linear in oscillator position. 

The nonlinear mechanical response we study can be described by a fully classical and noise-free theory. However, we also consider the effects of both thermal and quantum fluctuations. In the presence of noise, the stable oscillation amplitudes are only metastable, since there is a possibility of switching from one stable state to another via thermal or quantum activation \cite{Dykman2007PRE}, or quantum tunneling. We consider the regime of weak single-photon optomechanical coupling and sufficiently low temperature such that this type of switching is mostly negligible, which enables us to study small fluctuations around a single stable state. However, noise-induced switching is necessarily relevant close to bifurcation points where one of the stable solutions vanish. We estimate the switching rate close to such points. This is relevant for applications, since it determines how close to a bifurcation point the system can be considered stable for practical purposes.

We also analyze several applications that utilize the nonlinear response of the optomechanical system. A useful feature of the setup is that it combines bistable behavior with optical damping (or cooling) of noise. This enables phase-sensitive amplification of small resonant mechanical forces, where the system can latch onto a widely different stable state as a consequence of a small, temporary signal. The dynamical attractor's dependence on optical power also enables similar latch amplification of small optical signals and thereby switching of optical beams with smaller optical signals. We also show that the dynamical system can be used as an optically controlled mechanical memory.

The article is composed as follows: In Section \ref{sec:Model}, we present the model and analyze the coherent response of the optomechanical system, as well as fluctuations and noise-induced switching between stable states. Section \ref{sec:MechAmpl} describes how, and to which degree, the setup can be used to detect and amplify small, resonant mechanical forces. In Section \ref{sec:OptAmpl}, we show how switching between stable states can be induced by optical signals for amplification and memory purposes. Section \ref{eq:ResTwoMode} presents an alternative but more complicated optomechanical setup which displays the same effects, but where the required mechanical amplitudes are smaller. Final remarks are presented in Section \ref{sec:Disc}.

\section{Model}
\label{sec:Model}

\subsection{Setup}

We consider an optical cavity mode coupled to a mechanical oscillator with position operator $x$. The coupling is described by the standard radiation pressure interaction $H_\mathrm{int} = \hbar g_0 x \, a^\dagger a$, where $g_0$ is the single-photon optomechanical coupling rate and $a$ is the photon annihilation operator in the frame rotating at the cavity resonance (angular) frequency $\omega_r$. The position operator $x =  (c + c^\dagger)$ is expressed by the phonon annihilation operator $c$ and in units of the zero point motion $x_\mathrm{zpf} = \sqrt{\hbar/(2m\omega_m)}$, where the effective mass is denoted $m$ and the mechanical resonance frequency is $\omega_m \ll \omega_r$. We let the cavity mode be coherently driven at the frequency $\omega_{r} - \omega_m$. We also include a mechanical drive at the oscillator's resonance frequency. 

\begin{figure}[ht]
\includegraphics[width=0.69\textwidth]{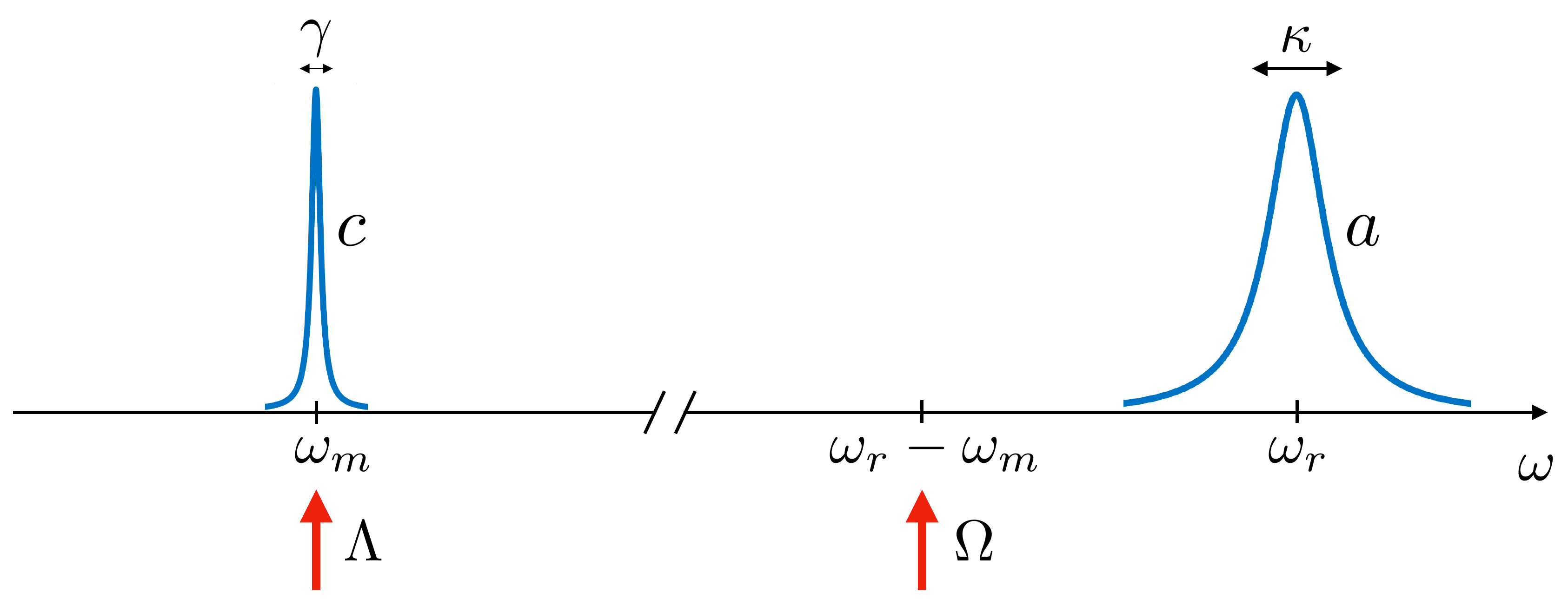}
\caption{Overview of the model parameters. The cavity mode with linewidth $\kappa$ is driven at a frequency red-detuned by $\omega_m$ from its resonance frequency $\omega_r$. The mechanical mode with intrinsic linewidth $\gamma$ is driven at its resonance frequency $\omega_m$. The red arrows represent the coherent drives, parametrized by the rates $\Omega$ and $\Lambda$. \label{fig:freqs}}
\end{figure}

The dynamics of the system is determined by the quantum Langevin equations
\begin{align}
\label{eq:OptLangevin}
\dot{a} & = -\frac{\kappa}{2} a - i g_0 x a + e^{i \omega_m t} \Omega + \sqrt{\kappa} \, \xi \\
\dot{c} & = -\left(\frac{\gamma}{2} + i \omega_m \right) c - i g_0 a^\dagger a + e^{-i \omega_m t} 
\Lambda + \sqrt{\gamma} \, \eta
\label{eq:MechLangevin} 
\end{align}
where $\kappa$ ($\gamma$) is the energy decay rate of the cavity (oscillator) and we have assumed that the effective mechanical oscillator quality factor is always large. The optical and mechanical drive strengths are denoted $\Omega$ and $\Lambda$, respectively. These rates are proportional to the square root of drive power and they can be considered real and positive without loss of generality. See Figure \ref{fig:freqs} for an overview of the model parameters.
The operator $\xi(t)$ describes quantum vacuum noise from coupling to external fields and the standard Markovian treatment \cite{Gardiner1985PRA,Clerk2010RMP} gives Gaussian noise with $\langle \xi(t) \xi^\dagger(t') \rangle = \delta(t-t')$ and $\langle \xi^\dagger(t) \xi(t') \rangle = \langle \xi(t) \xi(t') \rangle = 0$, assuming that the temperature $T$ obeys $k_B T \ll \hbar \omega_c$.  We let the mechanical noise operator $\eta$ obey $\langle \eta(t) \eta^\dagger(t') \rangle = (n_\mathrm{th} + 1) \delta(t-t')$, $\langle \eta^\dagger(t) \eta(t') \rangle = n_\mathrm{th}$, and $\langle \eta(t) \eta(t') \rangle = 0$, with the thermal phonon number defined as
\begin{equation}
\label{eq:nthDef}
n_\mathrm{th} = \frac{1}{\mathrm{exp}[\hbar \omega_m/(k_B T)] - 1} \ . 
\end{equation}
We do not consider technical noise in any of the drives, but the theory can easily be extended to include that.

The discussion below will be limited to a particular regime of optical drive powers. The combination of coherent mechanical oscillations and optical power in the cavity gives rise to radiation pressure forces on the mechanical oscillator at all multiples of the mechanical frequency. This includes a DC force which shifts the equilibrium position of the oscillator which in turn shifts the average resonance frequency of the cavity. However, this frequency shift is negligible if we limit ourselves to low enough powers such that 
\begin{equation}
\label{eq:Restr}
 \frac{g_0^2 |a_\mathrm{max}|^2 }{\omega_m \kappa}  \ll 1 \ ,
\end{equation}
where $a_\mathrm{max} = 2\Omega/\kappa$ is the cavity amplitude one would get if the optical drive was on cavity resonance and $g_0$ was zero. In the limit \eqref{eq:Restr}, and for an oscillator with a large quality factor, we can also neglect the response of the mechanical oscillator at higher multiples of the mechanical resonance frequency and only take into account its response at or around its resonance frequency $\omega_m$. We consider this regime throughout this article.

We emphasize that the drive parametrized by $\Lambda$ can be realized through optical driving without the need for mechanical actuation. This can be achieved simply by adding an additional optical drive which is modulated at $\omega_m$ and thus gives rise to a coherent force on the mechanical oscillator, as was for example done in the study reported in Ref.~\cite{Krause2015PRL}. However, in Appendix \ref{app:Lambda}, we show that to realize the effects described in this article, we would require that this modulated beam drives an auxiliary optical cavity mode whose single-photon coupling to the mechanical oscillator is much smaller than that of cavity mode $a$. This ensures that a sufficiently large value of $\Lambda$ can be realized by optical means without significantly influencing the decay rate $\gamma$ or the mechanical noise properties defined above. 

\subsection{Coherent response}

\begin{figure}[t]
\includegraphics[width=0.69\textwidth]{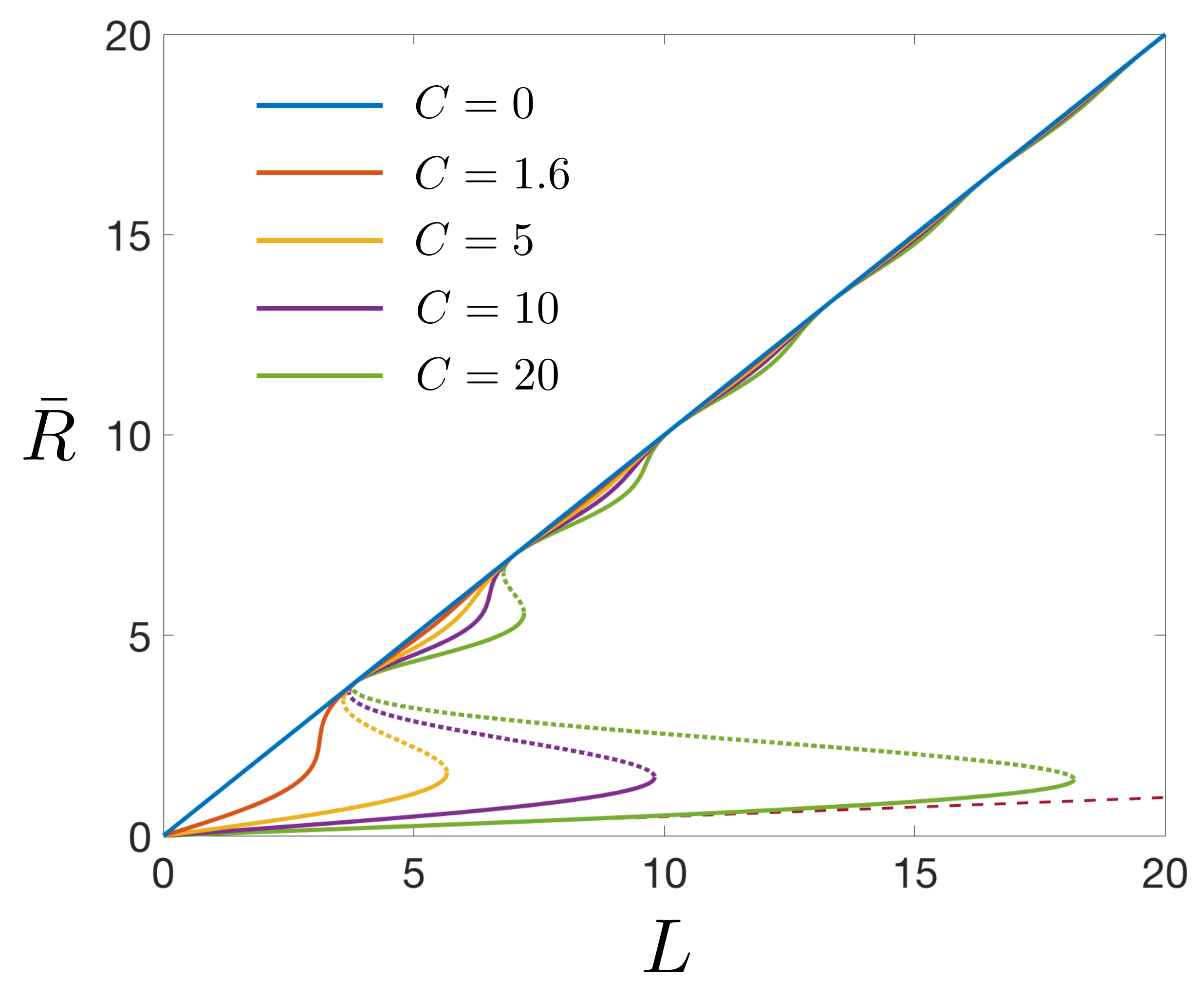}
\caption{The steady-state amplitude $\bar{R}$ of the coherent mechanical oscillations in the presence of a resonant mechanical drive with strength $L$ for $\omega_m/\kappa = 10$. The cavity mode is driven by a single beam red-detuned by one mechanical frequency $\omega_m$. The cooperativity $C$ is determined by the optical drive strength $\Omega$ according to Equation \eqref{eq:CoopMechDrive}. The dashed red line is the mechanical response at $C = 20$ that one would find from the linearized equations of motion.  \label{fig:NonlinResSM}}
\end{figure}

We begin by determining the amplitude and phase of the coherent mechanical oscillations, as well as the coherence of the optical cavity field. Initially, we neglect the thermal and quantum noise, but we will return to the role of fluctuations in Section \ref{sec:fluct}. As mentioned above, we neglect the mechanical oscillator's response at higher multiples of $\omega_m$. The coherent motion of the mechanical oscillator can then be expressed as 
\begin{equation}
\label{eq:xdefClass}
x(t) = x_0 + \bar{r} \cos(\omega_m t - \bar{\phi}) \ ,
\end{equation}
where $x_0$ is a shift of the equilibrium position, $\bar{r}$ is the oscillation amplitude and $\bar{\phi}$ is a phase. For convenience, we define the rescaled mechanical amplitude
\begin{equation}
\label{eq:Xdef}
\bar{R} = \frac{g_0 \bar{r}}{\omega_m}  \ .
\end{equation}
This allows us to express the cavity field as 
\begin{equation}
\label{eq:aClassdef}
a(t) = \sum_k e^{-ik (\omega_m t - \bar{\phi})} a_k  \quad , \quad  a_k = (-1)^{k-1} \Omega \, e^{i \bar{\phi}} \sum_{n}  \chi_n  J_{n-k}(\bar{R})  J_{n+1}(\bar{R}) \ , 
\end{equation}
where the sums over integers go from minus infinity to plus infinity, we have defined the susceptibilities
\begin{equation}
\label{eq:chiDef}
\chi_n = \frac{1}{\kappa/2 - i n \omega_m} \ ,
\end{equation}
and $J_n(\bar{R})$ is the $n$-th order Bessel function of the first kind. The quantity $a_k$ describes the optical coherence at frequency $\omega_r + k \omega_m$. We now define the standard optomechanical cooperativity
\begin{equation}
\label{eq:CoopMechDrive}
C = \frac{4 g_0^2 \alpha^2}{\kappa \gamma}
\end{equation}
where $\alpha = |\chi_{-1}| \Omega$ would be the intracavity coherent amplitude if $g_0 = 0$. The average mechanical amplitude $\bar{r}$ and phase $\bar{\phi}$ can then be determined by the complex, nonlinear, algebraic equation
\begin{equation}
\label{eq:XEqGen}
\bar{R} + C \sum_{n}  H_n  J_{n+1}(\bar{R}) J_{n}(\bar{R}) - e^{-i \bar{\phi}} L = 0 \ ,
\end{equation}
which follows from Equation \eqref{eq:MechLangevin} when defining 
\begin{equation}
\label{eq:HnDef}
H_n =  - \frac{i\kappa \, \chi^\ast_{n-1} \chi_n }{\omega_m |\chi_{-1}|^2} \ 
\end{equation}
and the rescaled mechanical drive strength
\begin{equation}
\label{eq:LDef}
L = \frac{4 g_0 \Lambda}{\gamma \omega_m}  \ .
\end{equation}
We note that the only physical parameter $H_n$ depends on is the dimensionless sideband parameter $\omega_m/\kappa$.

\begin{figure}[ht]
\includegraphics[width=0.59\textwidth]{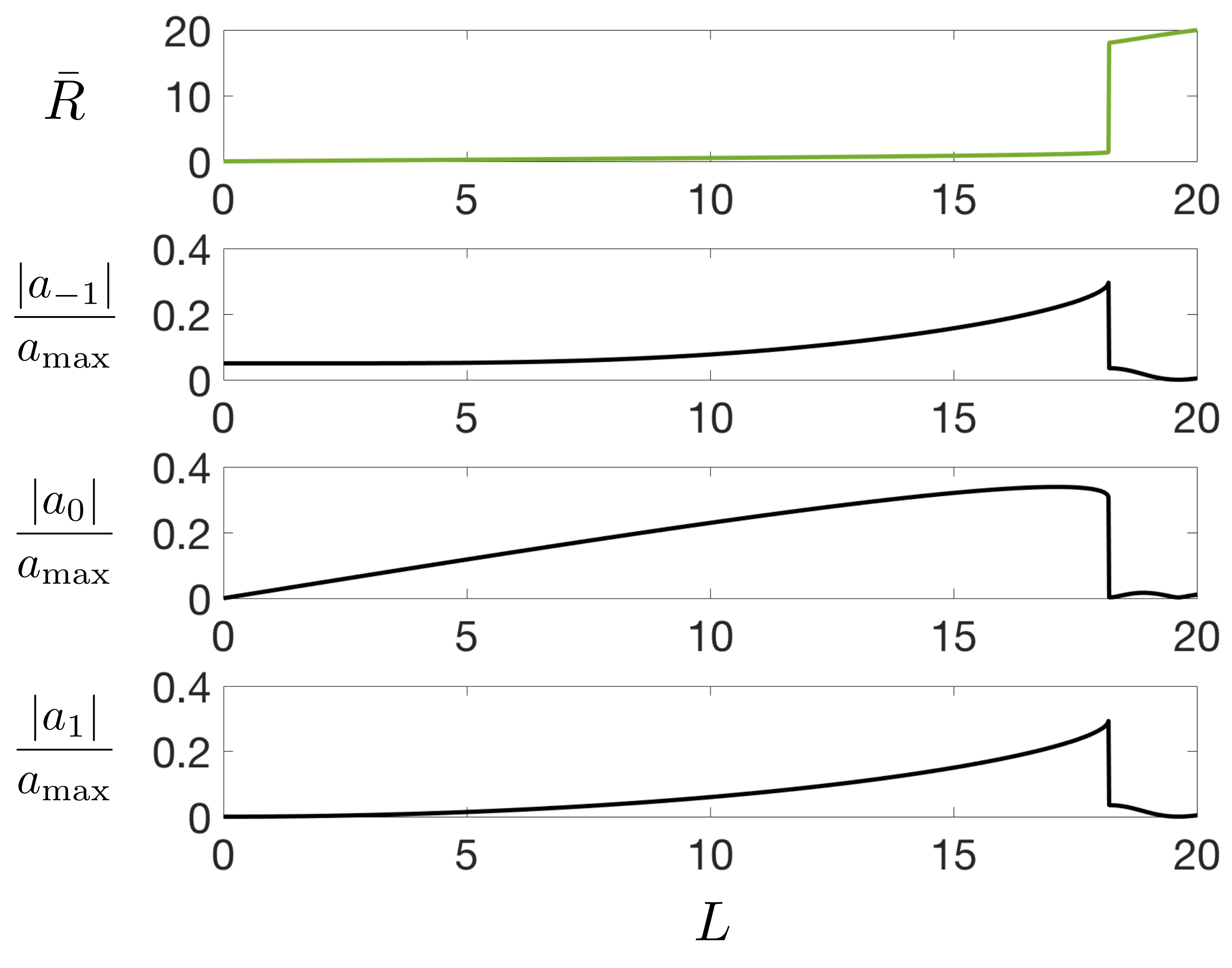}
\caption{Steady-state response of the optomechanical system when sweeping up the mechanical drive from $L = 0$ to $L = 20$ for cooperativity $C = 20$ and sideband parameter $\omega_m/\kappa = 10$. {\it Top panel:} Mechanical amplitude. {\it Lower three panels}: Optical coherence at frequencies $\omega_r - \omega_m$ (upper), $\omega_r$ (middle), and $\omega_r + \omega_m$ (lower) relative to $a_\mathrm{max} = 2\Omega/\kappa$. \label{fig:OptResponse}}
\end{figure}

We now consider the solutions to Equation \eqref{eq:XEqGen}. This equation can be solved numerically by truncating the sum at a sufficiently large integer. Let us focus on the mechanical response to the drive, i.e., on the mechanical amplitude $\bar{R}$ as a function of drive amplitude $L$, which is shown in Figure \ref{fig:NonlinResSM} for different values of the cooperativity $C$ and for a sideband parameter $\omega_m/\kappa = 10$.  We do not show the solution for the phase $\bar{\phi}$, but find that it is small ($ |\bar{\phi}|/\pi < 0.02$) for all cooperativities and drive strengths in Figure \ref{fig:NonlinResSM}. In fact, one can show that $\bar{\phi} = 0$ in the resolved sideband limit $\omega_m/\kappa \rightarrow \infty$. The Figure shows that for values of the scaled amplitude $\bar{R}$ much less than 1, the mechanical response is linear with an effective damping rate $\gamma (1+C) $. This is due the standard optomechanical damping (or cooling) mechanism, where the radiation pressure interaction leads to an enhanced mechanical decay rate \cite{Braginsky2002PhysLettA,Marquardt2007PRL,Wilson-Rae2007PRL}. However, when $\bar{R}$ becomes on the order of unity, the response becomes highly nonlinear. The physical explanation is that the amplitude of the cavity resonance frequency variations then become comparable to $\omega_m$, which is on the order of the detuning of the optical beam, such that the optical damping mechanism becomes inefficient. 

Increasing the cooperativity beyond a critical value $C_\mathrm{crit}$, Figure \ref{fig:NonlinResSM} shows that we get three solutions for $\bar{R}$ for an interval of drive strengths $L_- < L < L_+$. We will refer to $L_-, L_+$ as bifurcation points or turning points. As we will show in Section \ref{sec:Potential}, one of the solutions (indicated by a dashed line) is unstable. The system thus displays bistability where two stable solutions exist for the same drive strength $L$. In general, the value of $C_\mathrm{crit}$ depends on the sideband parameter $\omega_m/\kappa$, but is independent of it in the limit $\omega_m/\kappa \gg 1$. In the example shown in Figure \ref{fig:NonlinResSM}, we find $C_\mathrm{crit} \approx 1.82$. We also see that for even larger values of the cooperativity $C$, the system can display even more solutions, i.e., multistability. The presence of more than one solution means that the system will display hysteresis, i.e, the steady-state mechanical amplitude for a given drive strength will depend on the history of the system. To connect with the theory of optomechanical self-sustained oscillations \cite{Marquardt2006PRL}, we reiterate that here we are addressing the dynamical attractor diagram in the red-detuned regime.

To explain the origin of bistability, let us look at the curve for $C = 20$ in Figure \ref{fig:NonlinResSM}. For large drives, e.g., $L \sim 15$, the lower stable solution where $\bar{R} \sim 1$ describes the case where the optical drive is efficiently driving the cavity mode, such that the mechanical oscillator is efficiently damped. The other stable solution, where $\bar{R} \sim L$, describes the situation where the mechanical amplitude and thus the cavity's resonance frequency variation is already so large that the optical drive is unable to efficiently address the cavity. The optical damping is then almost absent and only the intrinsic mechanical damping remains. We note that the vanishing of the lower stable solution occurs when the effective frictional force on the oscillator becomes negative. This is because photon down-conversion processes where phonons are created can dominate for large oscillation amplitudes and a sufficient amount of cavity photons. Multistability with more than two stable solutions comes from the fact that for a region of amplitudes, e.g., around $\bar{R} \sim 5$ in Figure \ref{fig:NonlinResSM}, the dominant damping mechanism is optical but of a nonlinear nature that corresponds to multi-phonon annihilation processes in the quantum formalism.

There will be optical coherence not only at the original laser frequency, but at all multiples of the mechanical frequency, according to Equation \eqref{eq:aClassdef}. The coherence amplitudes $|a_k|$ depend on the steady-state mechanical amplitude $\bar{R}$. For $C > C_\mathrm{crit}$, the amplitude will jump from $\bar{R} \sim 1$ to a significantly larger value as the drive strength $L$ is sweeped up from 0 and beyond the bifurcation point where the lower stable solution vanishes. This jump will be accompanied by jumps in the optical coherences $a_k$ and can thus easily be monitored optically. The behaviour of optical coherences as $L$ is sweeped up is shown in Figure \ref{fig:OptResponse} for $k = {-1 , 0 , 1}$, $C = 20$, and $\omega_m/\kappa = 10$. We note that the coherences at other values of $k$ are significantly smaller in the resolved sideband limit $\omega_m/\kappa \gg 1$.

\subsection{Effective potential}
\label{sec:Potential}

\begin{figure}[t]
\includegraphics[width=0.89\textwidth]{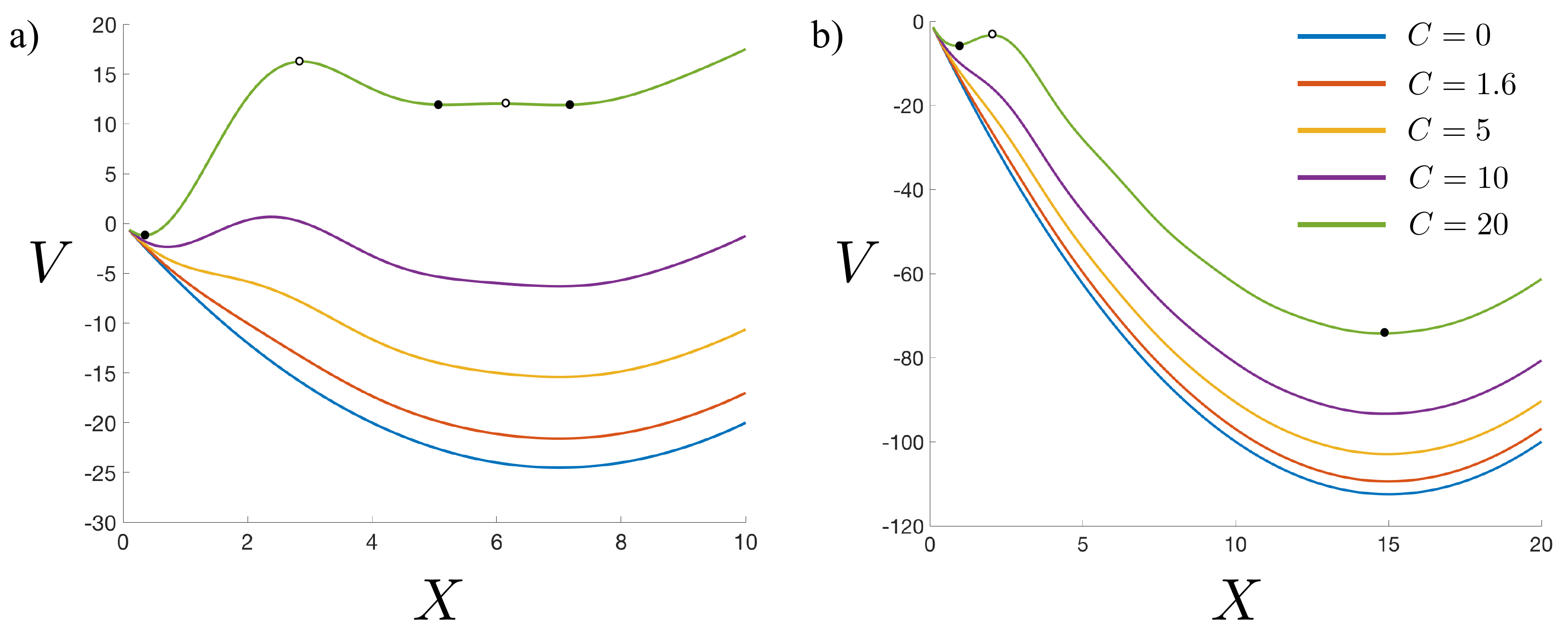}
\caption{The effective potential $V(X,0)$, as defined in Equation \eqref{eq:PotDef} for a) $L = 7$ and b) $L = 15$, with $\omega_m/\kappa = 10$. The filled (open) circles indicate stable (unstable) points (shown only for $C = 20$). \label{fig:PotMechDrive}}
\end{figure}

We will now show that, in the resolved sideband limit $\omega_m/\kappa \gg 1$, the steady-state solutions for the amplitude $\bar{R}$ can be interpreted as the positions of extremal points of an effective potential. This will be a helpful picture to have in mind when we later discuss fluctuations. Let us now consider the mechanical amplitude $r(t)$ and phase $\phi(t)$ defined in Equation \eqref{eq:xdefClass} as dynamical variables, even though we still ignore noise. We can equivalently describe the mechanical oscillations with rescaled quadrature variables $X(t)$ and $Y(t)$, defined by
\begin{equation}
\label{eq:QuadDef}
\frac{g_0 }{\omega_m} r(t) \cos (\omega_m t - \phi(t)) = X(t) \cos(\omega_m t)  + Y(t) \sin(\omega_m t)  \ .
\end{equation}
We then define the vector
\begin{equation}
\label{eq:QuadDefRescaled}
\mathbf{R} = (X,Y)  \ ,
\end{equation}
and the dimensionless time variable
\begin{equation}
\label{eq:TDef}
\tau =  \frac{\gamma t}{2}    \ .
\end{equation}
This allows us to write the equations of motion for the mechanical quadratures as
\begin{equation}
\label{eq:RVecEOM}
\frac{d\mathbf{R}}{d\tau} = \mathbf{F}(\mathbf{R})   
\end{equation}
where 
\begin{align}
F_X(\mathbf{R})  & = - X - C \, \mathrm{Re} \left[e^{i \phi} \sum_{n}  H_n J_{n}(R) J_{n+1} (R) \right] + L \label{eq:FX} \\
F_Y(\mathbf{R})  & = - Y - C \, \mathrm{Im} \left[e^{i \phi} \sum_{n}  H_n J_{n}(R) J_{n+1} (R) \right] \label{eq:FY} 
\end{align}
and $R e^{i\phi} \equiv X+iY$. We can interpret Equation \eqref{eq:RVecEOM} as determining the position $\mathbf{R}$ of a fictitious particle in two dimensions subject to a (dimensionless) force $\mathbf{F}(\mathbf{R})$ in the limit of large friction. Note that if we require the time derivative in Equation \eqref{eq:RVecEOM} to vanish and let $R \rightarrow \bar{R}$, these equations reduce to the complex Equation \eqref{eq:XEqGen}. While the vector field $\mathbf{F}(\mathbf{R})$ is not generally conservative, we can, in the resolved sideband limit $\omega_m/\kappa \gg 1$, write $\mathbf{F}(\mathbf{R}) \approx - \nabla V(\mathbf{R})$ with
\begin{equation}
\label{eq:PotDef}
V(\mathbf{R}) = \frac{1}{2}R^2 + 2 C \left\{1 - \left[J_0(R)\right]^2 - \left[J_1(R)\right]^2 \right\} - LX \ .
\end{equation}
This follows from the fact that $H_n \approx 1$ for $n \in \{0,1\}$ and $H_n \approx -i \kappa/(n(n-1) \omega_m)$ for $n \notin \{0 , 1\}$ in the limit $\omega_m/\kappa \gg 1$. We note that in the limit of small amplitudes, $R \ll 1$, we get $V(\mathbf{R}) = (1+C)R^2/2 - LX$. This limit thus gives a displaced, quadratic potential stiffened by the factor $1+C$ compared to the case without optical drive, in accordance with standard linearization of the optomechanical interaction.

The locations of the minima (maxima) of the potential $V(\mathbf{R})$ then correspond to stable (unstable) steady-state positions for the fictitious particle, and hence to stable (unstable) values for the quadrature variables $X$ and $Y$. The potential is shown in Figure \ref{fig:PotMechDrive} for the same cooperativities as in Figure \ref{fig:NonlinResSM} and for two different values of the drive strength $L$. We have set $Y = 0$ since the stable points lie on the $X$-axis in the resolved sideband limit. We see that for $C = 20$ and $L = 7$, we have three minima, whereas for $C = 20$ and $L = 15$, we have two minima. This is in accordance with the exact numerical solutions shown in Figure \ref{fig:NonlinResSM}.

\subsection{Fluctuations around a stable solution}
\label{sec:fluct}

We now take into account the noise acting on both the optical field and the mechanical oscillator, which is represented mathematically in Equations \eqref{eq:OptLangevin} and \eqref{eq:MechLangevin} by the operators $\xi(t)$ and $\eta(t)$. The steady-state solutions for the mechanical amplitude $\bar{R}$ are then only metastable, since thermal or quantum noise can cause switching between them. However, we will focus on the regime of thermal phonon number $n_\mathrm{th} \ll (\omega_m/g_0)^2$ and weak single-photon coupling $g_0 \ll \kappa , \omega_m$ and we will show in Section \ref{sec:Switch} that the switching rates are then negligibly small except for drive strengths $\Lambda$ close to the bifurcation points. This means that for most values of $\Lambda$, we can focus on fluctuations around a single, stable equilibrium.

Let us consider small fluctuations around one of the equilibrium values of $\bar{R}$ and $\bar{\phi}$. We write
\begin{align}
\label{eq:aWithFluct}
a (t) & = e^{-i \bar{R} \sin(\omega_m t - \bar{\phi})} \left( \bar{a}(t) + d(t) \right) \\
c(t) & = \frac{x_0}{2} + e^{-i (\omega_m t - \bar{\phi})} \left(\frac{\bar{r}}{2}  + \tilde{c}(t) \right)
\end{align}
defining the quantities 
\begin{equation}
\label{eq:aBarDef}
\bar{a}(t) = \sum_k e^{-i k (\omega_m t- \bar{\phi})} \, \bar{a}_k   \quad , \quad \bar{a}_k =   \chi_{k} \Omega \, e^{i\phi}  J_{-(k+1)}(\bar{R}) \ 
\end{equation}
for convenience. Inserting this into Equations \eqref{eq:OptLangevin} and \eqref{eq:MechLangevin} and neglecting nonlinear terms in the fluctuation variables $d$ and $\tilde{c}$ gives
\begin{align}
\label{eq:OptLangevinFluct}
\dot{d} & = -\frac{\kappa}{2} d  - ig_0 \bar{a} \left(e^{-i (\omega_m t - \bar{\phi})} \tilde{c} + e^{i (\omega_m t - \bar{\phi})} \tilde{c}^\dagger \right)+  \sqrt{\kappa} \, \tilde{\xi} \\
\dot{\tilde{c}} & = - \frac{\gamma}{2} \tilde{c} - i g_0 e^{i (\omega_m t - \bar{\phi})} \left(\bar{a}^\ast d + \bar{a} \, d^\dagger \right) +\sqrt{\gamma} \tilde{\eta} \ ,
\label{eq:MechLangevinFluct}
\end{align}
where $\tilde{\xi}$ ($\tilde{\eta}$) is $\xi$ ($\eta$) multiplied by a time-dependent phase factor and thus obeys the same correlation properties as $\xi$ ($\eta$). We may adiabatically eliminate the cavity field fluctuations $d(t)$ if the mechanical oscillator dynamics (in the frame rotating at $\omega_m$) are slow compared to the cavity decay time $\sim 1/\kappa$. Upon assuming this, we find
 \begin{equation}
\label{eq:dSol}
d(t) = \zeta(t) - i \sum_n G_n \left(e^{-i (n+1) (\omega_m t -  \phi)} \chi_{n+1} \tilde{c}(t) +  e^{-i (n-1) (\omega_m t - \phi)} \chi_{n-1} \tilde{c}^\dagger(t) \right) 
\end{equation}
with $G_n = g_0 \bar{a}_n $ and 
\begin{equation}
\label{eq:zetaDef}
\zeta(t) =  \sqrt{\kappa} \int_{-\infty}^t  ds \,  e^{-\frac{\kappa}{2}(t - s)} \tilde{\xi}(s) \ .
\end{equation}
In general, the time-dependent coefficients in Equations \eqref{eq:OptLangevinFluct} and \eqref{eq:MechLangevinFluct} preclude a steady state solution to expectation values. However, if we assume, as before, that the mechanical oscillator only responds at or near the resonance frequency $\omega_m$ and not at other multiples of $\omega_m$, we can find approximate steady-state expressions for the Gaussian mechanical noise. By defining
\begin{align}
\label{eq:GammaPsiDef}
\Gamma & = \gamma + 2 \sum_{n} |G_n|^2 \left(\chi_{n+1} - \chi_{n-1}^\ast \right)  \ ,  \\
\Psi & = 2 \sum_n G^\ast_{n-2} G_n \left(\chi_{n-1} - \chi_{n-1}^\ast \right) \ ,
\end{align}
we arrive at an equation of motion for $\tilde{c}$ alone,
\begin{equation}
\label{eq:cTildeAfterAd}
\dot{\tilde{c}} = - \frac{\Gamma}{2} \tilde{c} - \frac{\Psi}{2} \tilde{c}^\dagger + \sqrt{\gamma} \, \tilde{\eta} - i \sum_n \left(G^\ast_{n-1}  \zeta_n + G_{n+1}  \zeta_n^\dagger \right) \ ,
\end{equation}
where we have defined
 \begin{equation}
\label{eq:zetanDef}
\zeta_n(t) = e^{i n (\omega_m t - \bar{\phi})} \zeta(t) \ .
\end{equation}

\begin{figure}[t]
\includegraphics[width=0.89\textwidth]{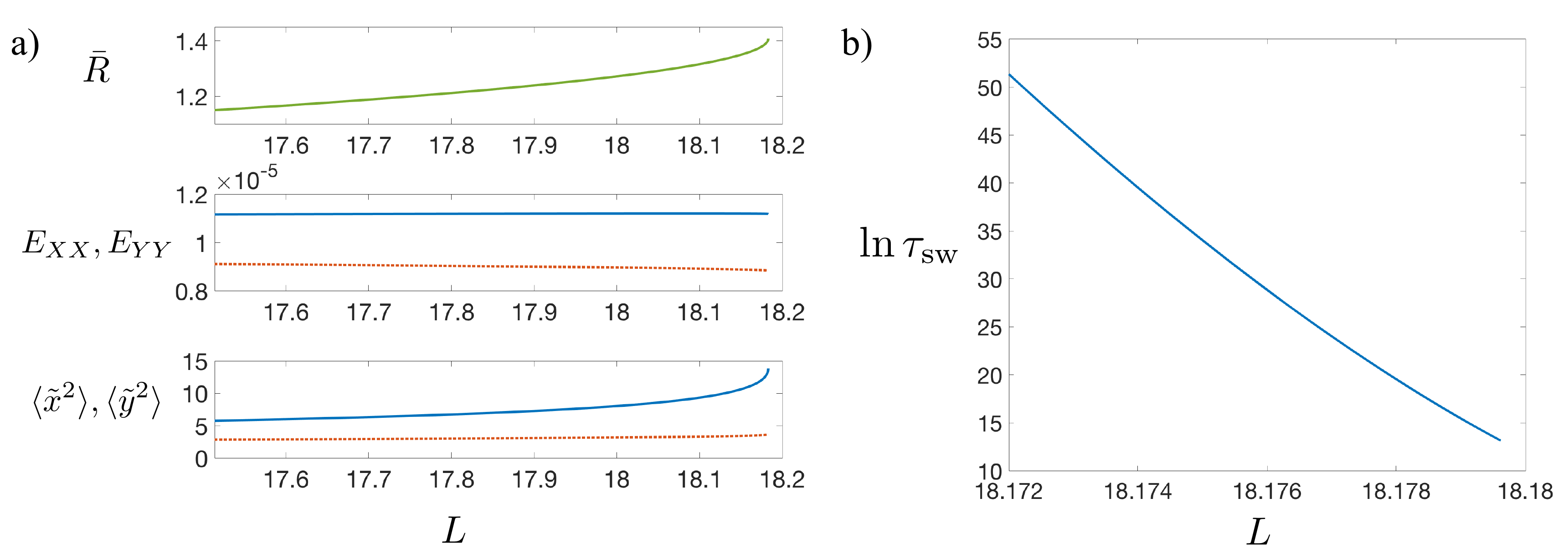}
\caption{Fluctuations and switching time for parameters $C = 20$, $\omega_m/\kappa = 10$, $g_0/\kappa = 0.005$, $\gamma/\kappa = 10^{-5}$, and $n_\mathrm{th} = 10$ when driving close to the bifurcation point $L_+$ ($\approx 18.183$ here). a) {\it Upper panel:} Steady-state amplitude $\bar{R}$. {\it Middle panel:} The noise variables $E_{XX}$ (solid/blue) and $E_{YY}$ (dashed/red) as defined in Equations \eqref{eq:noiseCorrelationsXX} and \eqref{eq:noiseCorrelationsYY}. {\it Lower panel:} Average size of quadrature fluctuations  $\langle \tilde{x}^2 \rangle$ (solid/blue) and $\langle \tilde{y}^2 \rangle$ (dashed/red) in units of the zero point motion $x_\mathrm{zpf}$. b) The logarithm of the dimensionless average switching time $\tau_\mathrm{sw}$ in Equation \eqref{eq:Kramers} (only plotted for $L$ such that $V(\mathbf{\bar{R}}_\mathrm{ust}) - V(\mathbf{\bar{R}}_\mathrm{st,l}) > 10 E$). Note the difference in scale on the $L$-axis between a) and b).   \label{fig:MechFluct}}
\end{figure}

We can now find expressions for expectation values involving the mechanical noise operator $\tilde{c}$. By once again using that the mechanical oscillator only responds near its resonance frequency, this leads to the average phonon occupancy (relative to the large coherent state)
\begin{equation}
\label{eq:PhononOcc}
\langle \tilde{c}^\dagger \tilde{c} \rangle = \frac{\gamma n_\mathrm{th}  + \mathrm{Re} \left(|\Psi|^2/(2\Gamma)\right) + \kappa \sum_n \left[|G_{n}|^2 + \mathrm{Re} \left(  G_{n-2}^\ast G_{n} \Psi^\ast / \Gamma \right)  \right] \left|\chi_{n-1}  \right|^2}{ \left(1 - |\Psi|^2/|\Gamma|^2 \right) \mathrm{Re} \, \Gamma} \ .
\end{equation}
This expression is more complicated than what one gets for an undriven, optically damped mechanical oscillator \cite{Marquardt2007PRL,Wilson-Rae2007PRL}, since there are nonzero optical coherences at more than one frequency. To understand this result, it is instructive to think about what one would find in the absence of mechanical driving and with an optical drive at a detuning $ l\omega_m$. We would then have $G_n = G \, \delta_{n,l}$, giving 
\begin{equation}
\label{eq:POSingleCoherence}
\langle \tilde{c}^\dagger \tilde{c} \rangle_{L = 0} = \frac{ \gamma n_\mathrm{th} + \kappa |G|^2 \left|\chi_{l-1} \right|^2}{\gamma + 2|G|^2 \, \mathrm{Re}(\chi_{l+1} - \chi^\ast_{l-1} )}   \ . 
\end{equation}
The second term in the denominator is the optical contribution to the damping rate, whereas the second term in the numerator describes the additional heating that comes from radiation pressure shot noise, or equivalently, from down-converting photons in Stokes processes. A naive generalization of this result to many beams would give the terms in \eqref{eq:PhononOcc} that do not contain $\Psi$. The presence of $\Psi$ is due to interference between the down-conversion of a photon from a detuning $n \omega_m$ to $(n-1) \omega_m$ and the up-conversion of a photon from a detuning $(n-2) \omega_m$ to $(n-1) \omega_m$. It is well-known that such interference effects can lead to nonzero correlations between orthogonal quadratures and even squeezing \cite{Kronwald2013PRA_2}. Mathematically, this follows from the nonzero expectation value
\begin{equation}
\label{eq:csq}
\langle  \tilde{c}^2 \rangle = -\frac{\Psi}{\Gamma} \left(\langle  \tilde{c}^\dagger \tilde{c} \rangle + \frac{1}{2} \right) - \frac{\kappa}{\Gamma} \sum_n  G_{n-2}^\ast G_{n} \left|\chi_{n-1}\right|^2  \ . 
\end{equation}

In Figure \ref{fig:MechFluct}a), we plot the fluctuations in the two quadratures
\begin{equation}
\label{eq:TildeDefs}
\tilde{x} = \tilde{c} + \tilde{c}^\dagger \quad , \quad \tilde{y} = i \left( \tilde{c}^\dagger - \tilde{c} \right)
\end{equation}
for cooperativity $C = 20$ and for drive strengths $L$ close to the bifurcation point $L_+$ where the lower stable solution vanishes. We observe that the noise in the $\tilde{x}$-quadrature grows rapidly when approaching the bifurcation point. The reason is the softening of the effective potential along this quadrature as the potential barrier is about to vanish. We emphasize that for the fluctuation analysis above to be valid, we must at least demand that $\langle \tilde{x}^2 \rangle , \langle \tilde{y}^2 \rangle \ll (\omega_m/g_0)^2$.

\subsection{Switching dynamics}
\label{sec:Switch}

We now address the possibility of switching from one equilibrium to another in the presence of bistability. The analysis in the previous section assumes that the system fluctuates around just one of the stable solutions. Although switching is mostly negligible, it is necessarily relevant for drives $L$ sufficiently close to the bifurcation points $L_-$ and $L_+$. 

We seek to describe the switching dynamics from the lower stable solution $(\bar{R} \sim 1)$ to the higher stable solution $(\bar{R} \sim L)$ for a drive strength $L$ close to the critical value $L_+$ where the lower stable solution vanishes. As an illustration, consider the potential $V$ for $C = 20$ and $L = 15$ shown in Figure \ref{fig:PotMechDrive}b). We wish to find out how long, on average, a particle remains trapped in the small potential minimum at $X \sim 1$ in the presence of noise before escaping downhill to large values of $X$. It is then not possible to separate the calculation of the coherent motion from that of the noise as we did above. We still assume that we can adiabatically eliminate the cavity field and consider the full stochastic dynamics of the rescaled quadrature variables $X$ and $Y$, defined in Equation \eqref{eq:QuadDef}. The equations of motion are 
\begin{align}
\label{eq:XYEOMSwitchX}
\frac{dX}{d\tau} & = F_X(\mathbf{R})  + F_{X,\mathrm{noise}}(\mathbf{R},\tau)  \\ 
\label{eq:XYEOMSwitchY}
\frac{dY}{d\tau} & = F_Y(\mathbf{R})  + F_{Y,\mathrm{noise}}(\mathbf{R},\tau) 
\end{align}
where $F_X$ and $F_Y$ were defined in Equations \eqref{eq:FX} and \eqref{eq:FY}, and
\begin{align}
\label{eq:FXnoise}
 F_{X,\mathrm{noise}}(\mathbf{R},\tau)  =  \frac{2 g_0}{\gamma \omega_m} \left\{\sqrt{\gamma}(\tilde{\eta} + \tilde{\eta}^\dagger) - i \sum_n \left[(G^\ast_{n-1}(\mathbf{R}) - G^\ast_{n+1}(\mathbf{R}) )  \zeta_n - (G_{n-1}(\mathbf{R}) - G_{n+1}(\mathbf{R}) ) \zeta_n^\dagger \right] \right\}   \\
\label{eq:FYnoise}
 F_{Y,\mathrm{noise}}(\mathbf{R},\tau)  =  \frac{2 g_0}{\gamma \omega_m} \left\{i \sqrt{\gamma}(\tilde{\eta}^\dagger - \tilde{\eta}) -  \sum_n \left[(G^\ast_{n-1}(\mathbf{R}) + G^\ast_{n+1}(\mathbf{R}) )  \zeta_n + (G_{n-1}(\mathbf{R}) + G_{n+1}(\mathbf{R}) ) \zeta_n^\dagger \right] \right\}  \ 
\end{align}
are Gaussian noise variables originating from the optical noise $\xi$ and the mechanical noise $\eta$. As before, we neglect the oscillator's response at higher multiples of its resonance frequency, which allows the noise correlation functions to be written
\begin{align}
\label{eq:noiseCorrelationsXX}
\langle  F_{X,\mathrm{noise}}(\tau) F_{X,\mathrm{noise}}(\tau') \rangle  & = \left(\frac{2g_0}{\omega_m} \right)^2 \left(n_\mathrm{th} + \frac{1}{2} + \frac{\kappa}{2\gamma} \sum_n |G_{n-1} - G_{n+1}|^2 |\chi_n|^2 \right) \delta(\tau - \tau') \equiv 2 \, E_{XX}  \, \delta(\tau - \tau') \\
\label{eq:noiseCorrelationsYY}
\langle  F_{Y,\mathrm{noise}}(\tau) F_{Y,\mathrm{noise}}(\tau') \rangle  & = \left(\frac{2g_0}{\omega_m} \right)^2 \left(n_\mathrm{th} + \frac{1}{2} + \frac{\kappa}{2\gamma} \sum_n |G_{n-1} + G_{n+1}|^2 |\chi_n|^2 \right) \delta(\tau - \tau')  \equiv 2 \, E_{YY}  \, \delta(\tau - \tau')  \\
\frac{1}{2} \langle  \left\{ F_{X,\mathrm{noise}}(\tau) \ , \ F_{Y,\mathrm{noise}}(\tau') \right\} \rangle  & = \left(\frac{2g_0}{\omega_m} \right)^2 \frac{\kappa}{\gamma} \sum_n \mathrm{Im} \left(G_{n-1} G^\ast_{n+1} \right) |\chi_n|^2  \delta(\tau - \tau')  \equiv 2 \, E_{XY}  \, \delta(\tau - \tau')  
\end{align}
where we have omitted the dependence on $\mathbf{R}$ for clarity. The properties $E_{XX} \neq E_{YY}$ and $E_{XY} \neq 0$ mean that the noise blob in phase space will be non-spherical and rotated with respect to the $X$- and $Y$-axes. We note, however, that in the resolved sideband regime $\omega_m/\kappa \gg 1$ and with our choice of phase for the mechanical drive, we have $E_{XX} \geq E_{YY}$ and $|E_{XY}| \ll E_{XX},E_{YY} $. 

We will now simplify the model given by Equations \eqref{eq:XYEOMSwitchX} and \eqref{eq:XYEOMSwitchY} in a way that still captures its essential features. This will allow us to estimate the average time before switching as well as to efficiently simulate the stochastic dynamics. We will consider the resolved sideband regime and use the {\it classical} stochastic equation
\begin{equation}
\label{eq:RVecEOMNOISE}
\frac{d\mathbf{R}}{d\tau} = -\nabla V(\mathbf{R})  + \sqrt{2 E} \, \mathbf{N}(\tau)
\end{equation}
to model the dynamics, where the potential $V$ is defined in Equation \eqref{eq:PotDef} and we assume isotropic noise satisfying 
\begin{equation}
\label{eq:NNOISEDef}
\langle N_X(\tau) N_X(\tau') \rangle = \langle N_Y(\tau) N_Y(\tau') \rangle = \delta(\tau - \tau') \quad , \quad  \langle N_X(\tau) N_Y(\tau') \rangle = 0 \ .
\end{equation}
The noise in our simplified model is quantified by $E \equiv E_{XX}(\mathbf{R} = \mathbf{\bar{R}}_\mathrm{st,l})$ where $\mathbf{\bar{R}}_\mathrm{st,l}$ is the lower stable solution to the noise-free equation of motion. In other words, $E$ is a measure of the thermal and quantum noise in the $X$-quadrature in the potential valley from which the fictitious particle eventually escapes. Note that this model will slightly overestimate the noise in the $Y$-quadrature, but this is not expected to influence our results in any significant way. The use of a classical model is justified by the fact that we have assumed a Gaussian state at all times. 

The average time $\tau_\mathrm{sw}$ before the fictitious particle escapes the potential minimum at $\mathbf{\bar{R}}_\mathrm{st,l}$ can now be estimated. The simplified model \eqref{eq:RVecEOMNOISE} determines the dissipative dynamics of a particle in a potential $V(\mathbf{R})$ subject to large friction and thermal fluctuations at a temperature proportional to $E$. A generalization of Kramer's escape rate \cite{Kramers1940Physica} to a two-dimensional potential \cite{Langer1969AnnPhys,Hanggi1986JStatPhys} then gives
\begin{equation}
\label{eq:Kramers}
\tau_\mathrm{sw} =  2 \pi \sqrt{\frac{\lambda_2(\mathbf{\bar{R}}_\mathrm{ust})}{|\lambda_1(\mathbf{\bar{R}}_\mathrm{ust})| \lambda_1(\mathbf{\bar{R}}_\mathrm{st,l}) \lambda_2(\mathbf{\bar{R}}_\mathrm{st,l})}} \times \mathrm{exp}\left[\frac{V(\mathbf{\bar{R}}_\mathrm{ust}) - V(\mathbf{\bar{R}}_\mathrm{st,l})}{E}\right] \ .
\end{equation}
Here, $\mathbf{\bar{R}}_\mathrm{ust}$ is the unstable steady-state solution, i.e., the position of the saddle point of the two-dimensional potential. We have also defined $\lambda_1$ and $\lambda_2$ as the eigenvalues of the Hessian matrix $H_{ij}(\mathbf{R}) = \partial_i \partial_j V(\mathbf{R})$, where $\lambda_1(\mathbf{\bar{R}}_\mathrm{ust})$ is the single negative eigenvalue at the saddle point. We note that the formula \eqref{eq:Kramers} is only valid when the exponent is much larger than 1. 

The average dimensionless time $\tau_\mathrm{sw}$ before switching is shown as a function of drive $L$ in Figure \ref{fig:MechFluct}b) for cooperativity $C = 20$. In this example, the critical drive strength is $L_+ \approx 18.183$. To relate to dimensionful parameters, $\tau_\mathrm{sw} \gg 1$ means that the average time before switching is much larger than the intrinsic mechanical decay time $2/\gamma$. For the sensing applications discussed below, it is worth noting that for the parameters such as those used in Figure \ref{fig:MechFluct}, the drive strength $L$ can be chosen very close to the critical value without risking accidental switching due to thermal fluctuations. However, we also emphasize that the time $\tau_\mathrm{sw}$ is exponentially dependent on the temperature $T \propto n_\mathrm{th}$.

\section{Phase-sensitive amplification of small resonant mechanical forces}
\label{sec:MechAmpl}

We now discuss a possibility for exploiting this setup to detect and amplify small mechanical forces oscillating at the resonance frequency $\omega_m$. We note that for a cooperativity $C$ just below the critical $C_\mathrm{crit}$, the system can be used as a phase-sensitive linear amplifier. While this can also be of interest for applications, we will focus on nonlinear amplification mechanisms in this article. There can be several different ways of exploiting bistability for amplification, depending on whether one wants to implement an active readout of information at a particular time \cite{Siddiqi2004PRL,Siddiqi2005PRL} or a passive sensor that detects a pulsed signal arriving at an unknown time. We will have the latter situation in mind here. The nonlinear amplification mechanism we study has the advantage that the pulse will cause the system to latch onto a widely different steady state, such that it can be read out at any subsequent point in time and possibly with less stringent requirements on additional low-noise amplification than required by a linear amplifier.

We consider a situation where the mechanical drive strength is set to $\bar{L}$ close to the bifurcation point $L_+$ where the lower stable solution vanishes. However, we let $L_+ - \bar{L}$ be sufficiently large such that the system fluctuates around the lower stable solution ($\bar{R}_\mathrm{st,l} \sim 1$) for very long times. To be more precise, we want the average time $\tau_\mathrm{sw}$ before switching to the upper stable solution ($\bar{R}_\mathrm{st,u} \sim L$) to be so large that we can ignore that possibility for all practical purposes. In addition to the mechanical force we deliberately apply, we imagine that the mechanical oscillator is also briefly subjected to another small resonant force that we wish to detect. The total drive strength becomes $L(\tau) = \bar{L} + l(\tau)$, where $l(\tau)$ describes the complex drive amplitude variations due to the additional force. With $\bar{L}$ real, the drive amplitude becomes 
\begin{equation}
\label{eq:Lamplitude}
|L(\tau)| = \sqrt{\left(\bar{L} + \mathrm{Re}[l(\tau)]\right)^2 + \left(\mathrm{Im}[l(\tau)]\right)^2} .
\end{equation}
We see that depending on the complex phase of $l(\tau)$, the perturbation can lead to an increased total amplitude $|L(\tau)|$. For simplicity, we will from now on assume that $l(\tau)$ is real and positive, but we emphasize that we are describing a phase-sensitive amplifier that will be most sensitive to forces that are in phase with the deliberately applied force.

\begin{figure}[t]
\includegraphics[width=0.89\textwidth]{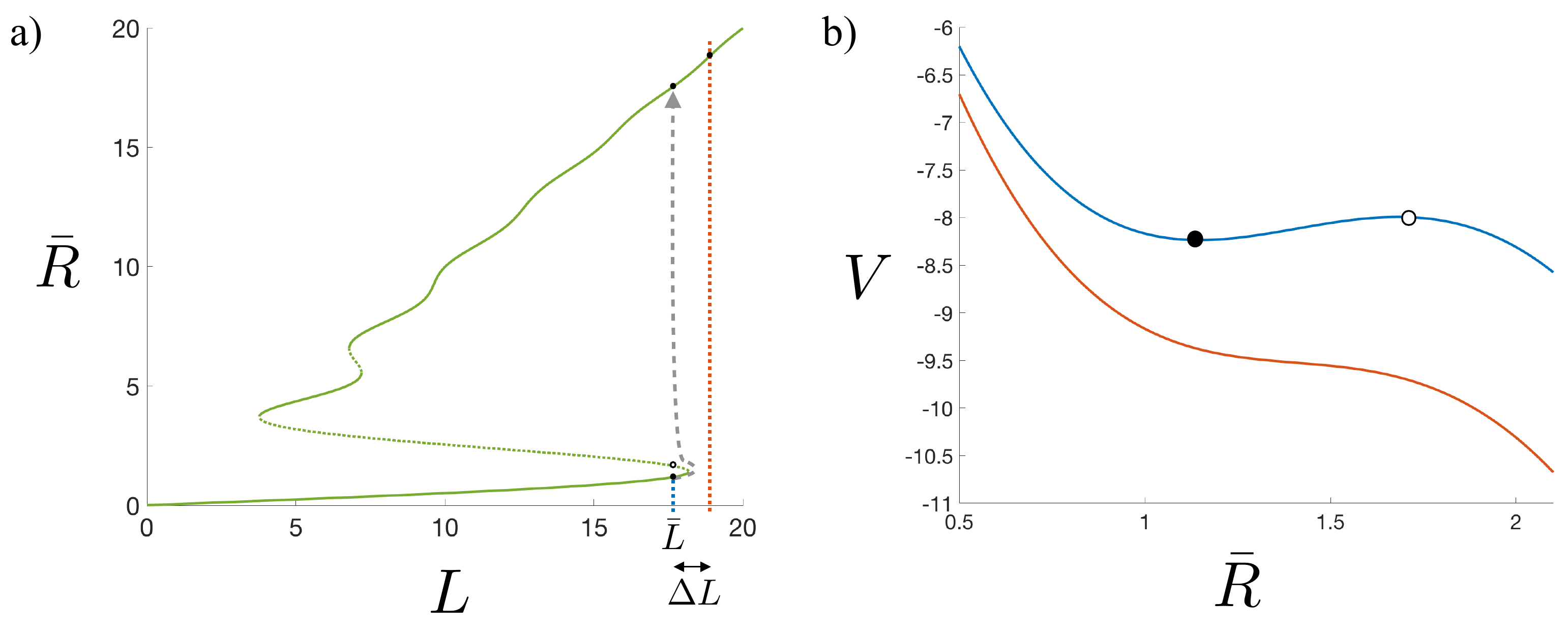}
\caption{Scheme for amplifying small mechanical signals. a) The steady-state response can change dramatically and permanently due to a small and temporary change in drive strength from $\bar{L}$ to $\bar{L} + \Delta L$. b) The effective potential has a stable minimum for small $\bar{R} \sim 1$ when $L = \bar{L}$ (upper/blue curve), but not for $L = \bar{L} + \Delta L$ (lower/red curve). For a pulse of size $\Delta L$ and of sufficient duration $\Delta T$, the steady-state amplitude can thus change permanently from $\bar{R}_\mathrm{st,l} \sim 1$ to $\bar{R}_\mathrm{st,u} \gg 1$. We have again used $C = 20$ and $\omega_m/\kappa = 10$. \label{fig:LatchMech}}
\end{figure}

Let us now discuss the amplification mechanism. We first assume that the duration of the pulse $l(\tau)$ is much longer than the intrinsic mechanical decay time $2/\gamma$. The system will then respond adiabatically to the change in drive strength. As shown in Figure \ref{fig:LatchMech}a), an increase in the total drive strength to a maximal value $\bar{L} + \Delta L$ can bring the system beyond the critical drive strength $L_+$ where the lower stable solution vanishes. In that case, the amplitude $R$ will increase towards the upper stable solution $\bar{R}_\mathrm{st,u} \sim L$. After the pulse $l(\tau)$ has passed, the drive strength returns to $\bar{L}$. The fictitious particle can then either return to the lower stable solution $\bar{R}_\mathrm{st,l} \sim 1$ or continue towards the upper stable solution $\bar{R}_\mathrm{st,u} \sim \bar{L}$, depending on which side of the potential maximum it finds itself. For a pulse of sufficient duration and strength, the system always ends up in the upper solution. This means that a small and temporary signal can cause a large and permanent change in the optomechanical system, which can easily be read out without the need for additional low-noise amplification.

The amplification mechanism can also work for pulses shorter than the mechanical decay time. The analysis of the amplification mechanism above assumed a long pulse, where we can think of the potential $V(\mathbf{R})$ temporarily changing from the blue (upper) curve to the red (lower) curve in Figure \ref{fig:LatchMech}b). In the case of a short pulse, however, it is better to think of the potential as fixed (the blue (upper) curve) and $l(\tau)$ as additional noise that can kick the fictitious particle over the barrier. This means that the amplitude of $l(\tau)$ must far exceed the thermal/quantum noise $\sim \sqrt{E}$. For a long pulse, on the other hand, where the potential changes adiabatically with the pulse, we will see that the amplification mechanism can be efficient for amplitudes $|l(\tau)|$ comparable to the existing broadband noise. 

We now quantify the amplification in the proposed latching scheme. To this end, we define the average input mechanical amplitude as $\bar{c}_\mathrm{in}(L) =  \Lambda/\sqrt{\gamma}$ and the average output amplitude as $\bar{c}_\mathrm{out}(\Lambda) = \sqrt{\gamma} \, \bar{r}(\Lambda)/2 -  \bar{c}_\mathrm{in}(\Lambda)$ for a single input/output port. The squares of the input and the output amplitudes give the incoming and outgoing phonon flux, respectively \cite{Gardiner1985PRA,Clerk2010RMP}. We define an amplifier power gain by 
\begin{equation}
\label{eq:GMechDef}
{\cal G}_m = \left(\frac{\bar{c}_\mathrm{out,u}(\bar{\Lambda}) -\bar{c}_\mathrm{out,l}(\bar{\Lambda})}{\bar{c}_\mathrm{in}(\bar{\Lambda}+\Delta \Lambda) -\bar{c}_\mathrm{in}(\bar{\Lambda})} \right)^2 \ .
\end{equation}
where $\bar{\Lambda}$ ($\Delta \Lambda$) relates to $\bar{L}$ ($\Delta L$) as in Equation \eqref{eq:LDef}. The numerator in \eqref{eq:GMechDef} is the square of the permanent difference in average output amplitude between the final state (after the latching mechanism has been triggered) and the initial state. The denominator is the square of the temporary difference in average input amplitude. Inserting the definition of the input and output amplitudes gives
\begin{equation}
\label{eq:GainMech}
{\cal G}_m =  \left(\frac{\bar{R}_\mathrm{st,u}(\bar{L}) - \bar{R}_\mathrm{st,l}(\bar{L})}{ \Delta L / 2 }\right)^2 \ .
\end{equation}
Another interpretation of the gain ${\cal G}_m$ is now apparent - up to a constant, it is the permanent amplitude change due to the pulse divided by the temporary amplitude change the pulse would cause in the absence of optomechanical coupling. For $C, \bar{L} \gg 1$, we can roughly write $\bar{R}_\mathrm{st,u}(\bar{L}) \approx \bar{L}$ and $\bar{R}_\mathrm{st,l}(\bar{L}) \sim 1 \ll \bar{L}$. The gain can then be approximated by
\begin{equation}
\label{eq:GainMech2}
{\cal G}_m = \left(\frac{2 \bar{L}}{\Delta L} \right)^2 \ .
\end{equation}
in the limits $C \gg 1$ and $\Delta L  , 1 \ll \bar{L}$. We reiterate that the minimum drive change $\Delta L$ one can detect is set both by the thermal and quantum noise as well as the duration of the pulse. 
\begin{figure}[t]
\includegraphics[width=0.79\textwidth]{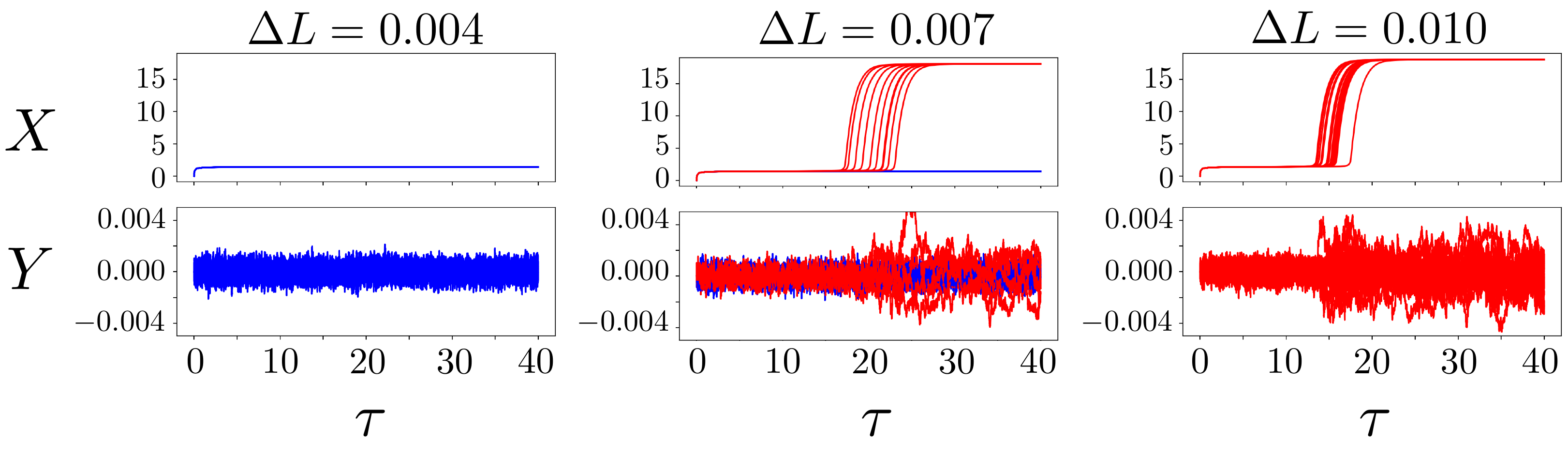}
\caption{Results from the classical stochastic simulations described in the text. The mechanical signal leads to a temporary change in drive strength according to the Gaussian \eqref{eq:Lpulse} centered at $\bar{\tau} = 10$. The results in this figure are for $\Delta \tau = 10$, i.e., for a pulse duration much longer than the intrinsic mechanical decay time. 20 trajectories are shown for three different values of $\Delta L$. Other parameters are $C = 20$, $\omega_m/\kappa = 10$, $n_\mathrm{th} = 10$, $g_0/\kappa = 0.005$, and $\gamma/\kappa = 10^{-5}$. The unperturbed drive strength is $\bar{L} = 18.179$, which means that $\tau_\mathrm{sw} \sim e^{15}$ and thus switching due to thermal/quantum fluctuations can safely be neglected. Trajectories that ended up at large amplitudes $R \sim \bar{L}$ have been colored red, and trajectories that remained at amplitudes $R \sim 1$ have been colored blue.  \label{fig:TimeTrace10dL}}
\end{figure}
\begin{figure}[h]
\includegraphics[width=0.49\textwidth]{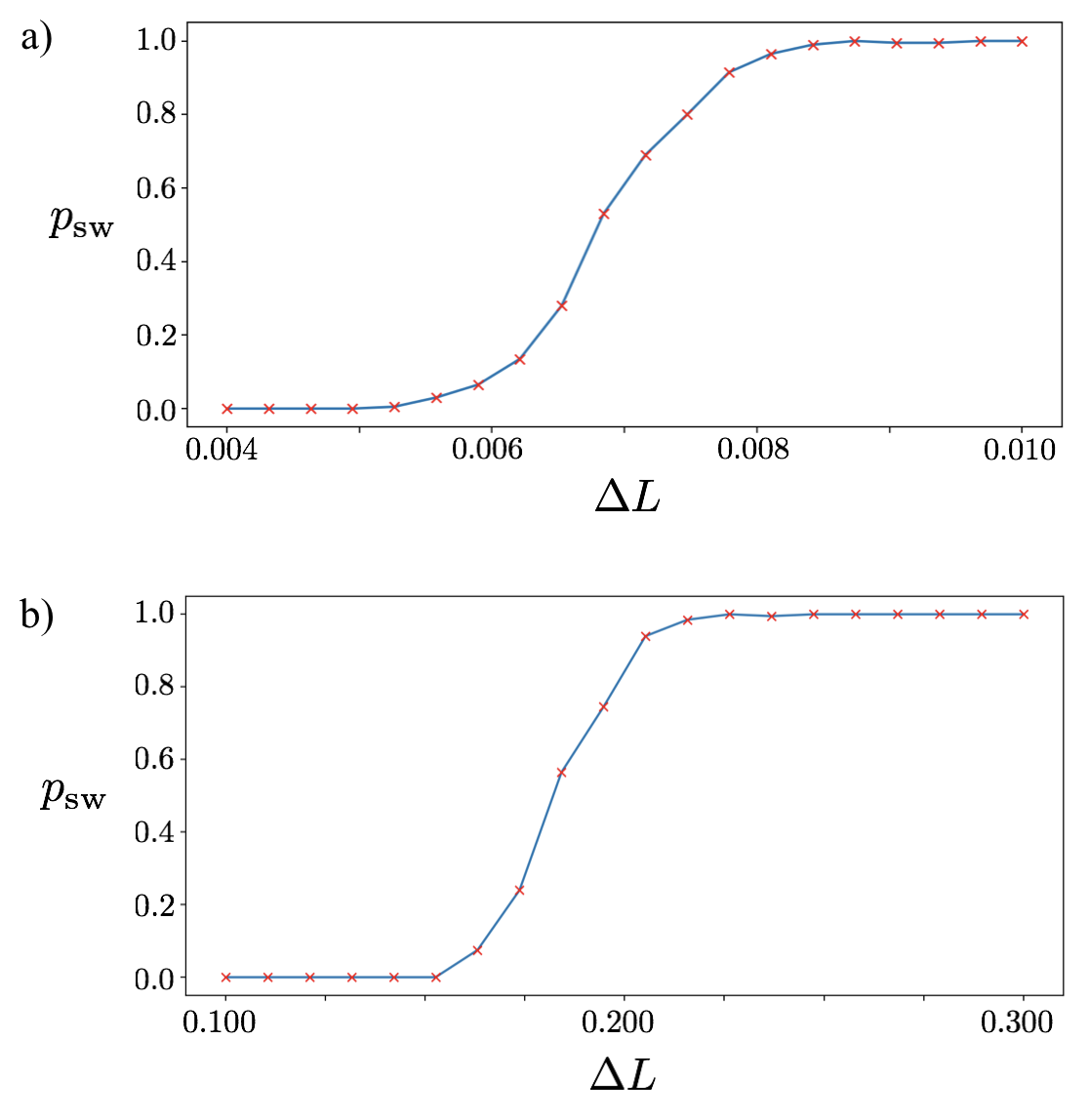}
\caption{Probability of switching $p_\mathrm{sw}$ from steady-state amplitude $\bar{R} \sim 1$ to $\bar{R} \sim {L}$ as a result of the mechanical pulse \eqref{eq:Lpulse}. The probabilities have been determined from 200 trajectories at each value $\Delta L$, limiting the precision to the percent level. The pulse duration is a) $\Delta \tau = 10$, b) $\Delta \tau = 0.1$. Other parameters are $C = 20$, $\omega_m/\kappa = 10$, $n_\mathrm{th} = 10$, $g_0/\kappa = 0.005$, and $\gamma/\kappa = 10^{-5}$, and the unperturbed drive strength is $\bar{L} = 18.179$.   \label{fig:ProbSwitchdL}}
\end{figure}

To go beyond the qualitative discussion above and demonstrate the amplifying mechanism, we have solved the classical, stochastic equation \eqref{eq:RVecEOMNOISE} numerically using the Euler-Maruyama method \cite{Kloeden1999Book}. The in-phase signal has been modeled as a Gaussian pulse 
\begin{equation}
\label{eq:Lpulse}
l(\tau) =  \Delta L \times \mathrm{exp}\left[- \frac{(\tau - \bar{\tau})^2}{2 (\Delta \tau)^2}\right] \ .
\end{equation}
The pulse has a maximum value $\Delta L$, it is centered at time $\bar{\tau}$, and it has a temporal width $\Delta \tau$. In Figure \ref{fig:TimeTrace10dL}, we show 20 time traces (or trajectories) of the quadratures $X$ and $Y$ for three different values of $\Delta L$, with $\bar{\tau} = 10$ and $\Delta \tau = 10$. We have again used cooperativity $C = 20$. The drive strength has been set to $\bar{L} = 18.179$, which according to Figure \ref{fig:MechFluct}b) means that switching due to noise is completely negligible on any reasonable time scale. To produce the time traces, the parameter $E$ has been chosen so as to agree with the values of $E_{XX}$ calculated in Figure \ref{fig:MechFluct}a). However, in the simulation we have modified Equation \eqref{eq:RVecEOMNOISE} such that we set $2E = (2g_0/\omega_m)^2 (n_\mathrm{th} + 1/2)$ when the amplitude $R = \sqrt{X^2 + Y^2}$ exceeds $0.8 L$. This simply reflects the fact that the noise around the upper stable solution is of mostly mechanical origin, unlike at the lower stable solution where it also originates from incoherent photon down-conversion. This modification has no significant influence on the switching dynamics.

The simulation results can be used to estimate the probability of switching $p_\mathrm{sw}$ to see how it depends on the pulse strength $\Delta L$ and the pulse duration $\Delta \tau$. The trajectories that switched to the upper solution have been colored red in Figure \ref{fig:TimeTrace10dL}, whereas the ones that did not switch are colored blue. We see that none of the trajectories switch to the upper stable solution when $\Delta L = 0.004$, some of them switch when $\Delta L = 0.007$, and that all of them switch when $\Delta L = 0.010$. The fact that none of the trajectories switch for small enough $\Delta L$ is in accordance with the assertion that switching due to thermal/quantum noise is negligible for our choice of parameters. In Figure \ref{fig:ProbSwitchdL}a), we plot the probability of switching $p_\mathrm{sw}$ as a function of pulse strength $\Delta L$ and pulse width $\Delta \tau = 10$, calculated from 200 simulated trajectories. We see that for $\Delta L \gtrsim 0.008$, the probability to trigger the latching mechanism is very close to 1. Note that in absence of both drives, we would have $\langle X^2 \rangle^{1/2} = \langle Y^2 \rangle^{1/2} = \sqrt{2E} = (2 n_\mathrm{th} + 1)^{1/2} g_0/\omega_m = 0.0023$ for our choice of parameters. This means that the minimal $\Delta L$ we can detect is comparable in magnitude to the intrinsic thermal or quantum fluctuations of the oscillator. Figure \ref{fig:ProbSwitchdL}b) shows the same plot, but for a pulse width $\Delta \tau = 0.1$, i.e., much shorter than the intrinsic mechanical decay time. In this case, a significantly stronger pulse $\Delta L \gtrsim 0.22$ is needed in order to have a large probability of switching. However, we see that the latching mechanism can be useful also in the non-adiabatic regime.

Finally, we briefly note that in the parameter regime where there are more than two stable states, i.e., multistability, it is possible to arrange the system such that it jumps between steady state branches irrespective of whether the pulse signal increases or decreases the drive amplitude. In other words, one could realize a bi-directional bifurcation amplifier \cite{Dong2018NatComm}.

\section{Optical switching and amplification}
\label{sec:OptAmpl}

\subsection{Amplification of optical signals by latching measurements}

\begin{figure}[ht]
\includegraphics[width=0.89\textwidth]{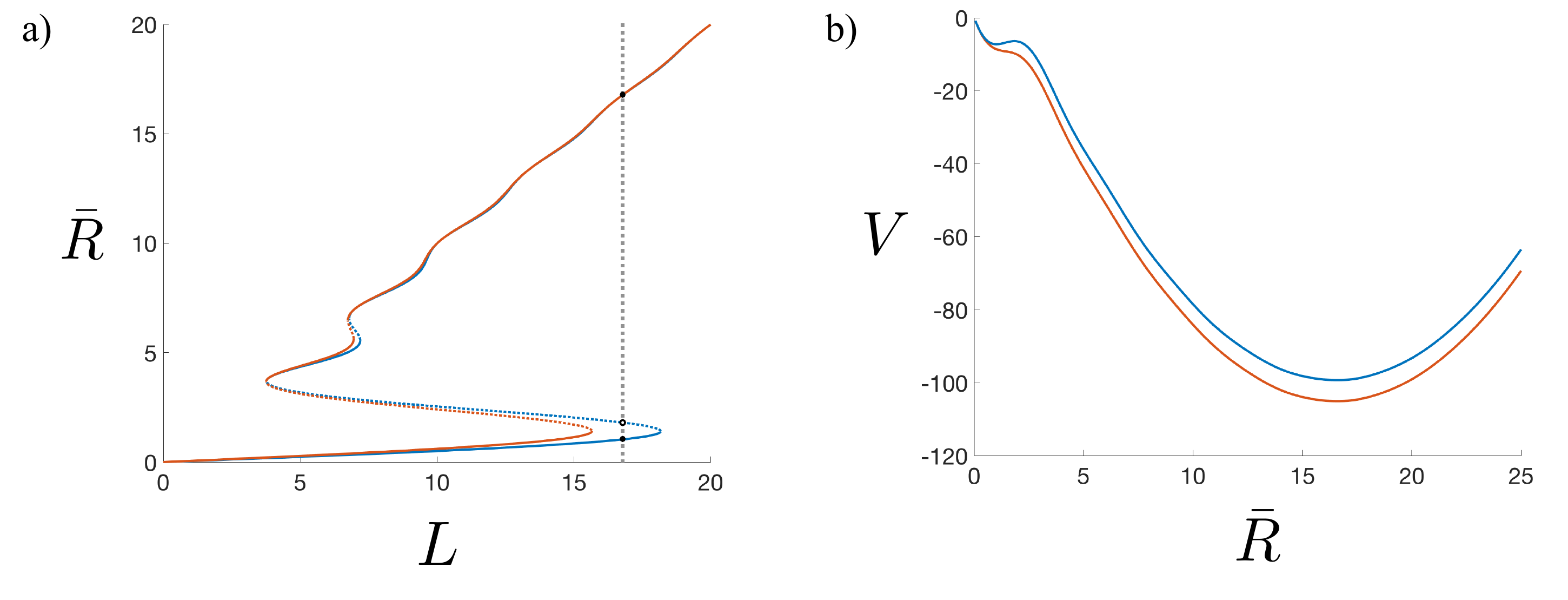}
\caption{Scheme for amplifying small optical signals. a) The steady-state response can change dramatically and permanently due to a small and temporary change in cooperativity from $\bar{C}$ to $\bar{C} - \Delta C$. b) The effective potential for $L$ indicated by the dashed line in a). The potential has a stable minimum for small $\bar{R} \sim 1$ when $C = \bar{C}$ (upper/blue curve), but not for $C = \bar{C} - \Delta C$ (lower/red curve). For a pulse of size $\Delta C$ and of sufficient duration $\Delta \tau$, the steady-state amplitude can thus change permanently from $\bar{R} \sim 1$ to $\bar{R} \sim L$. We have used $\bar{C} = 20$, $\Delta C = 3$, and $\omega_m/\kappa = 10$. Note that this choice of $\Delta C$ is exaggerated in order to demonstrate the mechanism. In practice, detection of far smaller $\Delta C/\bar{C}$ is feasible \label{fig:LatchOpt}}
\end{figure}
\begin{figure}[h]
\includegraphics[width=0.49\textwidth]{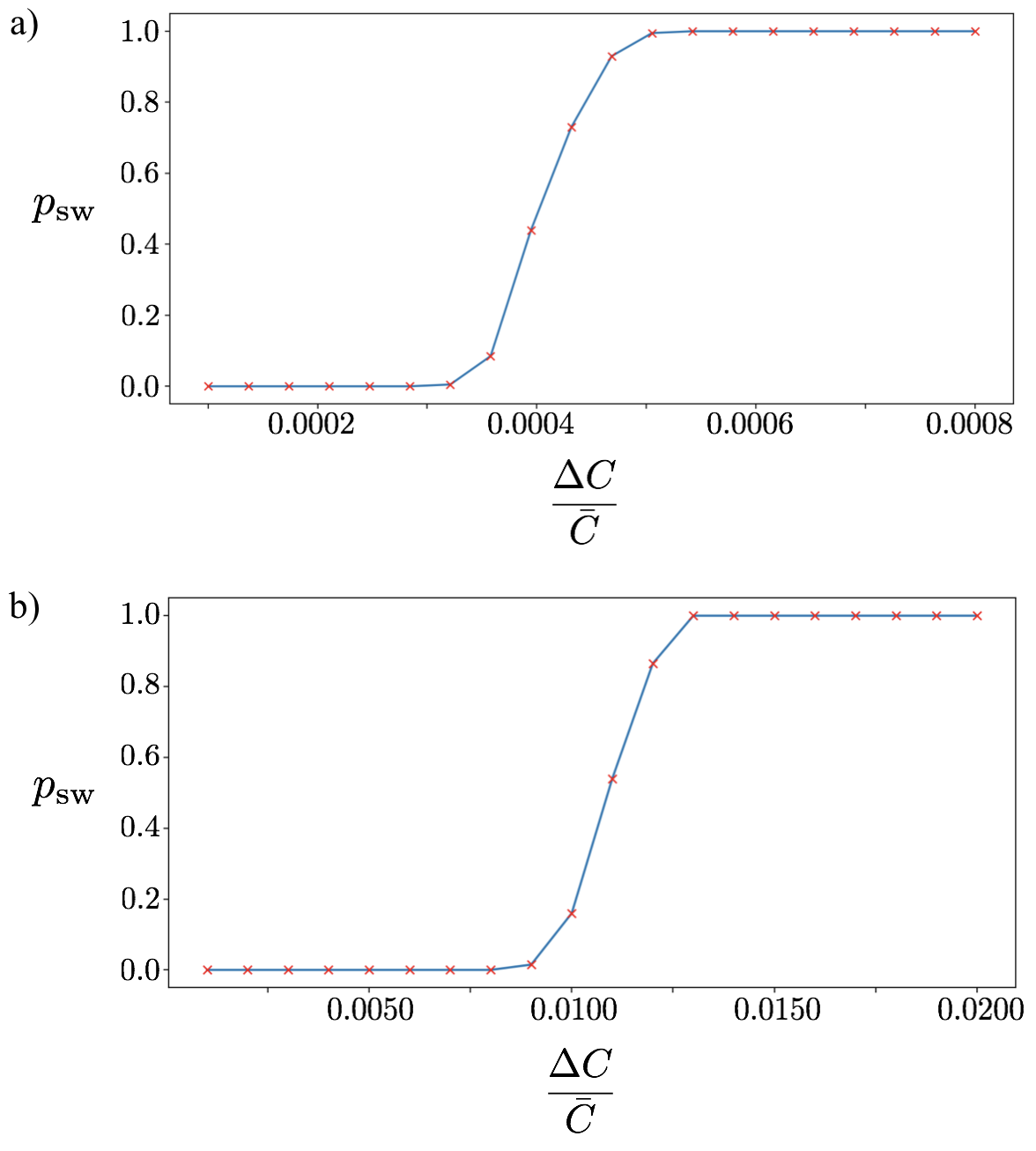}
\caption{Probability of switching from steady-state amplitude $\bar{R} \sim 1$ to $\bar{R} \sim {L}$ as a result of the optical pulse \eqref{eq:Cpulse}. The probabilities have been determined from 200 trajectories at each value $\Delta C$, limiting the precision to the percent level. The pulse duration is a) $\Delta \tau = 10$, b) $\Delta \tau = 0.1$, and the unperturbed cooperativity is $\bar{C} = 20$. Other parameters are $L = 18.179$, $\omega_m/\kappa = 10$, $n_\mathrm{th,eff} = 10$, $g_{0,a}/\kappa = 0.005$, and $\gamma/\kappa = 10^{-5}$. \label{fig:ProbSwitchdC}}
\end{figure}
The bistable response can also be used to detect and amplify small optical signals, which we now discuss. The mechanism is illustrated in Figure \ref{fig:LatchOpt}a). We consider a value of the cooperativity $\bar{C}$ and a fixed drive strength $L$ (dashed vertical line) below but close to the bifurcation point $L_+$. If the system resides in the lower stable state where $\bar{R} \sim 1$, a sufficiently large negative change in the cooperativity to $C(\tau) = \bar{C} - c(\tau)$ can decrease the critical drive strength $L_+$ such that $L > L_+$. This will force the mechanical amplitude to grow to $\bar{R} \sim L$, since the lower stable state vanishes. If the size and duration of the pulse $c(\tau)$ was sufficiently large, the system will eventually reside in the upper stable state even when the cooperativity returns to $\bar{C}$. This means that a small, negative change in optical power can lead to a large permanent change in the mechanical amplitude and thereby strongly influence the optical response of the cavity.

We now quantify the optical amplification in this scheme. The average input amplitude is defined as $\bar{a}_\mathrm{in}(\Omega)  =  \Omega/\sqrt{\kappa}$, which means that the square of the input amplitude is the incoming photon flux to the cavity. We consider first a constant change in cooperativity $C = \bar{C} - \Delta C$ caused by a change in the optical drive strength to $\Omega = \bar{\Omega} - \Delta \Omega$, such that $\Delta C/\bar{C} \approx 2 \Delta \Omega/\bar{\Omega}$ for $\Delta \Omega/\Omega \ll 1$. We denote the average output amplitude {\it at the cavity resonance frequency} $\bar{a}_{0,\mathrm{out}}(\Omega)  = \sqrt{\kappa} \,  a_0 (\Omega)$, where $a_0$ is defined according to Equation \eqref{eq:aClassdef}. The amplifier power gain is then defined by 
\begin{equation}
\label{eq:GOptDef}
{\cal G}_o = \left|\frac{\bar{a}_{0,\mathrm{out,l}}(\bar{\Omega})  -   \bar{a}_{0,\mathrm{out,u}}(\bar{\Omega}) }{  \bar{a}_\mathrm{in}(\bar{\Omega}+\Delta \Omega)  -  \bar{a}_\mathrm{in}(\bar{\Omega}) } \right|^2 
\end{equation}
Inserting for $a_0$ gives
\begin{equation}
\label{eq:GOpt2}
{\cal G}_o = \left(\frac{\Omega}{  \Delta \Omega } \right)^2 \left[\sum_n  \kappa \chi_n \left(J_n(\bar{R}_\mathrm{st,l}) J_{n+1}(\bar{R}_\mathrm{st,l}) - J_n(\bar{R}_\mathrm{st,u}) J_{n+1}(\bar{R}_\mathrm{st,u}) \right) \right]^2 
\end{equation}
In the resolved sideband regime, for large cooperativity and $L$ close to $L_+$, we have $\bar{R}_\mathrm{st,l} \sim 1$ and $J_n(\bar{R}_\mathrm{st,u}) \approx 0$, which gives
\begin{equation}
\label{eq:GOpt2}
{\cal G}_o \sim  {\cal O} \left(\frac{\bar{C}}{\Delta C}\right)^2 \ .
\end{equation}
We see that a large gain can be realized for a sufficiently small cooperativity change $\Delta C/\bar{C}$. However, it is again the thermal or quantum noise that limits how small $\Delta C$ can be.

To demonstrate the optical latch amplification mechanism, we have again solved the classical, stochastic equation \eqref{eq:RVecEOMNOISE} numerically as in Section \ref{sec:MechAmpl}. For simplicity, we model the cooperativity change as a Gaussian,
\begin{equation}
\label{eq:Cpulse}
c(\tau) =  \Delta C \times \mathrm{exp}\left[- \frac{(\tau - \bar{\tau})^2}{2 (\Delta \tau)^2}\right] \ ,
\end{equation}
with a maximum value $\Delta C$, centered at time $\bar{\tau}$, and temporal width $\Delta \tau$. We have again chosen a value of $L$ such that noise-induced switching is negligible for the unperturbed cooperaticity $\bar{C}$. The simulation results are similar to the ones in Section \ref{sec:MechAmpl}, showing that the probability of switching from the lower to the upper stable state rises with increasing $\Delta C$. In Figure \ref{fig:LatchOpt}, we show the switching probability as a function of $\Delta C/\bar{C}$ for two different pulse durations. For a long pulse $\Delta \tau = 10$ which the system can follow adiabatically, we find that the probability of switching is close to unity for $\Delta C/\bar{C} \gtrsim 5 \cdot 10^{-4}$. For a shorter pulse $\Delta \tau = 0.1$, we find that $\Delta C/\bar{C} \gtrsim 1.5 \cdot 10^{-2}$ is necessary for a large probability of switching due to the pulse.

\subsection{Mechanical memory}

\begin{figure}[h]
\includegraphics[width=0.49\textwidth]{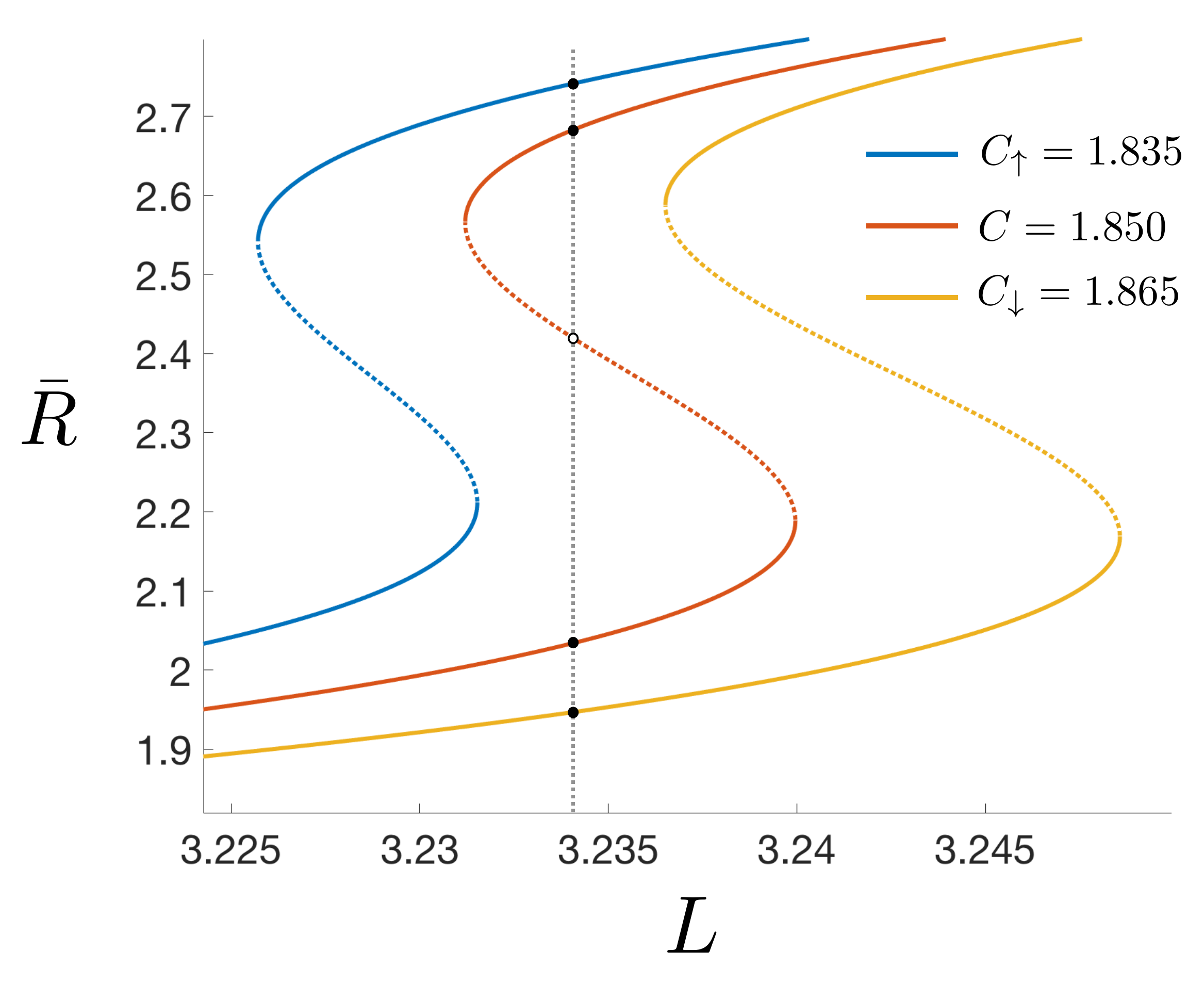}
\caption{Illustration of how the mechanical oscillation can be used as an optically controlled memory. The nominal cooperativity is $C$ (middle/red curve) and the drive strength is chosen in the region where the response is bistable. If the system is in the higher amplitude state ($\bar{R} \approx 2.68$), one can optically switch to the lower amplitude state ($\bar{R} \approx 2.04$) by temporarily changing the cooperativity to $C_\downarrow$ (lower/yellow curve) by adjusting the optical power. Likewise, switching from lower to higher steady-state amplitude can be done by temporarily changing the cooperativity to $C_\uparrow$ (upper/blue curve).    \label{fig:MechMem}}
\end{figure}
We will now briefly consider how the bistable behaviour can also be used to store binary data. The middle curve in Figure \ref{fig:MechMem}, shown in red, depicts the mechanical steady-state amplitude $\bar{R}$ as a function of drive strength $L$ for a cooperativity $C = 1.850$, which is just above the critical cooperativity $C_\mathrm{crit} \approx 1.82$. Let us assume that $L$ is fixed at the position of the vertical dotted line. If the system resides in the lower stable state, we can switch to the upper stable state by changing the cooperativity to $C_\uparrow < C$ by slightly adjusting the optical power. The value $C_\uparrow$ is chosen such that only the upper stable state remains for the fixed value of $L$, as shown by the upper (blue) curve in Figure \ref{fig:MechMem}. Once the amplitude has increased beyond a certain value, the system will end up in the upper stable state even when the cooperativity is returned to $C$. In other words, the system can be switched from the lower to the upper stable state by a small temporary decrease in optical input power. Conversely, if one wants to switch from the upper to the lower state, this can be done by a slight increase in optical input power resulting in a temporary increase in cooperativity to $C_\downarrow$ for which there is no upper stable state - see the lower (yellow) curve in Figure \ref{fig:MechMem}. This means that the mechanical memory can be switched back and forth simply by optical means.

This discussion is based on the system adiabatically following the change in cooperativity, which assumes that the optical pulses are much longer than the intrinsic mechanical decay time. As with the amplification scheme, it is possible to switch between the lower and the upper state with shorter pulses, but the reponse time of the memory, i.e., the time it takes to reach the other stable state, is set by the mechanical decay time.

\section{Bistability with resonant two-mode optomechanics}
\label{eq:ResTwoMode}

The phenomena we have described in this article rely on the possibility of realizing large coherent mechanical amplitudes $\bar{r} > \omega_m/g_0$. At the same time, we have assumed the hierarchy $g_0 \ll \kappa \ll \omega_m$. Such large amplitudes could pose practical challenges, for example due to intrinsic mechanical nonlinearities that might become relevant at large oscillation amplitudes. That being said, experimental efforts to explore the optomechanical attractor diagram on two different platforms do not seem to have encountered such problems \cite{Krause2015PRL,Buters2015PRA}. Nevertheless, in this Section, we will show that the same phenomena can be realized in so-called resonant two-mode optomechanics \cite{Heinrich2010PRA,Safavi-Naeini2011NJP,Ludwig2012PRL,Stannigel2012PRL}, but with a relaxed requirement on the size of the oscillation amplitudes. In this resonant case, we will see that the nonlinear response becomes relevant already when $\bar{r} \sim \kappa/g_0$.

\subsection{Setup}

\begin{figure}[t]
\includegraphics[width=0.89\textwidth]{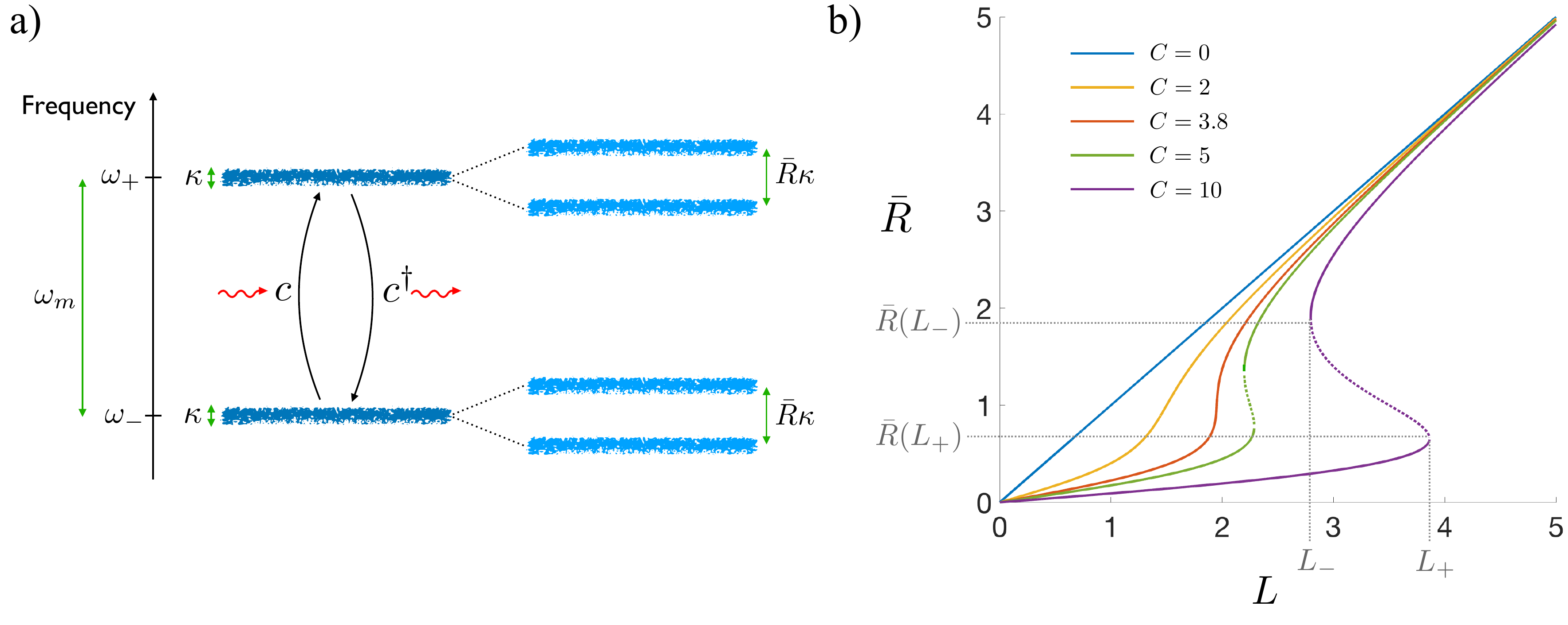}
\caption{Two-mode resonant optomechanics. a) Photons can resonantly scatter between two optical cavity modes with frequencies $\omega_\pm$ by the creation or annihilation of phonons. Coherent driving of the mechanical oscillator can lead to the equivalent of Autler-Townes splitting, where the size of the splitting is set by the mechanical amplitude. b) The nonlinear response to a mechanical drive. \label{fig:TwoModeRes}}
\end{figure}

We now define the model which includes two degenerate cavity modes $a_1$ and $a_2$ that are coupled by photon tunneling at a rate $J$ and that both couple to the same mechanical mode with the same rate, but opposite signs:
\begin{align}
\label{eq:HTwoModeRes0}
H =  \hbar J \left( a_1^\dagger a_2  + a_2^\dagger a_1 \right)  + \hbar g_0 \, x \left(a_1^\dagger a_1 - a_2^\dagger a_2  \right) + \hbar \omega_m c^\dagger c \ .
\end{align}
We note that the model \eqref{eq:HTwoModeRes} can be realized on several experimental platforms, for example the membrane-in-the-middle setup \cite{Thompson2008Nature,Ludwig2012PRL} or optomechanical crystals \cite{Eichenfield2009Nature_2,Safavi-Naeini2011NJP}.  For a tunneling rate $J$ exceeding the cavity linewidths, it is convenient to switch to a basis $a_+$ and $a_-$ in which the photon sector is diagonal. The modes $a_+$ and $a_-$ will then differ in frequency by $2J$. We assume that $J$ can be controlled such that this frequency difference matches the mechanical frequency $\omega_m$. The Hamiltonian can then be written
\begin{align}
\label{eq:HTwoModeRes}
H =   \hbar g_0 \left(\tilde{a}_+^\dagger \tilde{a}_- \tilde{c} + \tilde{a}_-^\dagger \tilde{a}_+ \tilde{c}^\dagger \right) \ ,
\end{align}
where the operators refer to frames rotating at the modes' respective resonance frequencies. The terms in the Hamiltonian describe processes where photons can scatter between the two cavity modes, either by creation or annihilation of a phonon. These processes are depicted in Figure \ref{fig:TwoModeRes}a). 
We have applied the rotating wave approximation by neglecting processes such as $a_+^\dagger a_-  c^\dagger $ that, at least from a naive viewpoint, do not conserve energy. We will comment on the validity of this approximation below.

\subsection{Coherent response}

Let us first find the average cavity amplitudes in the presence of both optical and mechanical drives. Including dissipation, noise, and drives gives the equations of motion:
\begin{align}
\label{eq:OptLangevinTwoModeResPlus}
\dot{\tilde{a}}_+ & = -\frac{\kappa}{2} \tilde{a}_+ - i g_{0} \, \tilde{c} \, \tilde{a}_- +  \sqrt{\kappa} \, \xi_+ \\
\label{eq:OptLangevinTwoModeResMinus}
\dot{\tilde{a}}_- & = -\frac{\kappa}{2} \tilde{a}_- - i g_{0} \, \tilde{c}^\dagger \, \tilde{a}_+ + \Omega + \sqrt{\kappa} \, \xi_- \\
\dot{\tilde{c}} & = - \frac{\gamma}{2}  \tilde{c} - i g_{0} \tilde{a}_-^\dagger \tilde{a}_+ + 
\Lambda + \sqrt{\gamma} \, \tilde{\eta} \ .
\label{eq:MechLangevinTwoModeRes} 
\end{align}
We have assumed that only the lower frequency cavity mode $\tilde{a}_-$ is coherently driven at resonance with strength parametrized by $\Omega$. For simplicity, we assume that the two cavities have the same linewidth $\kappa$, and that the cavity modes couple to independent vacuum noise fields $\xi_\pm$. The mechanical drive is parametrized by $\Lambda$ as before. Defining the rescaled amplitude
\begin{equation}
\label{eq:RDefTwoModeRes}
\bar{R} = \frac{g_0 \bar{r}}{\kappa}   \ ,
\end{equation} 
and ignoring noise, the classical, steady-state cavity amplitudes become
\begin{align}
\label{eq:aBarMinusTwoModeResMinusPlus}
\tilde{a}_- & = \frac{\alpha}{1 + \bar{R}^2} \quad , \quad  \tilde{a}_+  = -  i \bar{R} \tilde{a}_- 
\end{align}
where $\alpha = 2\Omega/\kappa$ would be the cavity amplitude in mode $\tilde{a}_-$ when $g_0 = 0$. We see that the amplitude $\tilde{a}_-$ becomes significantly reduced as the mechanical amplitude $\bar{R}$ becomes comparable to 1. Also, for $\bar{R} > 1$, we get population inversion where $|\tilde{a}_+| > |\tilde{a}_-|$. The decrease in cavity amplitude can be understood in terms of Autler-Townes splitting \cite{Autler1955PR,Heinrich2010PRA} known from atomic physics, as illustrated in Figure \ref{fig:TwoModeRes}a). When the mechanical oscillator has a large coherent amplitude and an average phase $\bar{\phi} = 0$, we can approximate the Hamiltonian by
\begin{align}
\label{eq:HTwoModeResAvg}
H = \hbar \frac{ \bar{R} \kappa }{2} \left(\tilde{a}_+^\dagger \tilde{a}_-  + \tilde{a}_-^\dagger \tilde{a}_+ \right) = \hbar   \frac{ \bar{R} \kappa }{2} \left(a_s^\dagger a_s - a_a^\dagger a_a \right) \ .
\end{align}
In the last equality, we have introduced the fields $a_{s (a)} = (\tilde{a}_+ \pm \tilde{a}_-)/\sqrt{2}$ to explicitly show that the mechanical driving induces a frequency splitting $\bar{R} \kappa$ between the effective cavity modes. This splitting becomes significant when it is comparable to the cavity linewidths, i.e., as $\bar{R}$ approaches unity which means $\bar{r} \rightarrow \kappa/g_0$.

The steady-state mechanical amplitude must be determined self-consistently from the equation
\begin{equation}
\label{eq:MechLangevinTwoModeR}
\bar{R} + \frac{C}{\left(1 + \bar{R}^2\right)^2} \bar{R} - L = 0 \ ,
\end{equation}
with the rescaled drive now defined as $L = 4g_0 \Lambda/(\kappa \gamma)$ and with the cooperativity
\begin{equation}
\label{eq:CTwoModeDef}
C =  \frac{4 g_0^2 \alpha^2}{\kappa \gamma}   
\end{equation} 
defined as before. In Equation \eqref{eq:MechLangevinTwoModeR}, we can clearly see that the optical damping term proportional to $C$ becomes less relevant as the amplitude $\bar{R}$ grows. In Figure \ref{fig:TwoModeRes}b), we plot the numerical solution to \eqref{eq:MechLangevinTwoModeR} for several different cooperativities. We see the same behaviour as in the single-mode case. Above a critical cooperativity, which is $C_\mathrm{crit} = 4$ in this case, the system displays bistability. From Equation \eqref{eq:MechLangevinTwoModeR} and in the limit $C \gg 1$, we can find analytic expressions for the bifurcation/turning points:
\begin{align}
\label{eq:LPlusMinTwoModeRes}
L_-  = \frac{4}{3} \sqrt[4]{3 C}  \quad ,  \quad L_+ & = \frac{3 \sqrt{3}}{16} C \ .
\end{align}
The corresponding amplitudes at these turning points are
\begin{align}
\label{eq:RLMinPlus}
\bar{R}(L_-)   = \sqrt[4]{3C} \quad , \quad \bar{R}(L_+)  & = \frac{1}{\sqrt{3}}  \ , 
\end{align}
also assuming $C \gg 1$. We emphasize that the same latching phenomena we discussed in the single-mode case can also be realized in this setup, but with a relaxed requirement on the size of the mechanical amplitude.

Finally, we briefly comment on the rotating wave approximation assumed in Equation \eqref{eq:HTwoModeRes}. From Figure \ref{fig:TwoModeRes}a), it is clear that this approximation cannot be valid for amplitudes such that the splitting $\bar{R} \kappa$ becomes comparable to $\omega_m$, i.e., when $\bar{r} \sim \omega_m/g_0$. From this, we may conclude that for $L < C$ and for cooperativities $C \ll \omega_m/\kappa$, the rotating wave approximation and the response shown in Figure \ref{fig:TwoModeRes}b) would be accurate. Conversely, for cooperativities $C > \omega_m/\kappa$, we must take into account the full model in order to determine the response at the upper stable state. In this case, it would be more useful to return to a description in terms of the original modes $a_1$ and $a_2$ as in Equation \eqref{eq:HTwoModeRes0} and think of the problem in terms of large-amplitude Landau-Zener-St\"{u}ckelberg oscillations \cite{Heinrich2010PRA}. We do not analyze this situation further here.

\section{Concluding remarks}
\label{sec:Disc}

We have investigated the nonlinear coherent response of an optically damped mechanical oscillator and showed that for sufficiently large optomechanical cooperativity, the system displays dynamical multistability. The analysis we have presented relates optical damping, known from standard linearized optomechanics, to the dynamical attractor diagram previously studied in connection with self-sustained oscillations. We have explored how a bistable dynamical response of an optomechanical system can be exploited in order to detect and amplify weak mechanical or optical signals. Comparing with the linear regime of optomechanical damping, the presented setup has the advantage that it can be biased at points in parameter space where the coherent response can be dramatically and permanently changed by small signals, while the thermal/quantum noise around the coherent response is optically damped (or cooled) at the same time. 

We have assumed a weak single-photon optomechanical coupling $g_0 \ll \kappa , \omega_m$ throughout this article. For large coupling rates $g_0 \sim \omega_m$, the multistable response would be smeared out by frequent switching between the equilibria. Even if this regime cannot be realized, it could be possible to see this kind of switching dynamics for drive strengths fine-tuned such that $C \sim C_\mathrm{crit}$ and $L \sim L_\pm$. In that case, one would realize non-Gaussian steady states with a potential for new applications.

\acknowledgments{
The author acknowledges useful discussions with Florian Marquardt and Aashish Clerk, and financial support from the Research Council of Norway through participation in the QuantERA ERA-NET Cofund in Quantum Technologies (project QuaSeRT) implemented within the European Union's Horizon 2020 Programme.
}

\appendix

\section{Optical realization of mechanical drive}
\label{app:Lambda}

In this Appendix, we will show that the mechanical drive parametrized by $\Lambda$ in Equation \eqref{eq:MechLangevin} can be implemented by additional optical driving. The additional drive could in principle address the same cavity mode $a$. However, we will see that to realize the effects discussed in this article, it will have to address an auxiliary cavity mode. 

We consider a model that includes two optical cavities with annihilation operators $a$ and $b$ that couple to the same mechanical oscillator. Cavity $a$ is driven by a single optical drive red-detuned by $\omega_m$ with drive strength $\Omega$, just as before. Cavity $b$ is driven at resonance by an optical drive with strength $\Omega_b$ that has been amplitude modulated at the mechanical frequency. For weak modulation, we only include the first order sidebands, which will address cavity $b$ with drive strength $\varepsilon \Omega_b \ll \Omega_b$ at detunings $\pm \omega_m$. The beat note between the carrier and the sidebands addressing cavity $b$ then leads to a coherent resonant force on the mechanical oscillator.

The equations of motion used to describe this system are
\begin{align}
\label{eq:OptLangevin2a}
\dot{a} & = -\frac{\kappa}{2} a - i g_{0} \, x \, a + e^{i \omega_m t} \Omega + \sqrt{\kappa} \, \xi \\
\label{eq:OptLangevin2b}
\dot{b} & = -\frac{\kappa}{2} b - i g_{0,b} \, x \, b + \left[1 + \varepsilon \left(e^{i \omega_m t} +  e^{-i \omega_m t}\right) \right] \Omega_b + \sqrt{\kappa} \, \xi_b \\
\dot{c} & = -\left(\frac{\gamma}{2} + i \omega_m \right) c - i g_{0} a^\dagger a  - i g_{0,b} b^\dagger b + \sqrt{\gamma} \, \eta \ ,
\label{eq:MechLangevin2} 
\end{align}
where $x = c + c^\dagger$ as before. For simplicity, we assume that the two cavities have the same linewidth $\kappa \ll \omega_m$. However, we allow for differing single-photon optomechanical coupling rates $g_{0} \neq g_{0,b} \ll \kappa$. We are still interested in the regime where the coherent amplitude of the oscillator $\bar{r}$ is so large that $g_{0} \bar{r}$ can exceed $\omega_m$. We will, however, assume that cavity mode $a$ is much more strongly coupled to the oscillator than cavity mode $b$, i.e., $g_{0} \gg g_{0,b}$. More precisely, we restrict ourselves to mechanical amplitudes such that 
\begin{equation}
\label{eq:Rrestriction}
\frac{g_{0,b} \bar{r}}{\omega_m} \ll \sqrt{\frac{\kappa}{\omega_m}} < 1 \ . 
\end{equation}
We also define the cooperativity associated with mode $b$:
\begin{equation}
\label{eq:Coopb}
C_{b} = \frac{4 g_{0,b}^2 (\chi_0 \Omega_{b})^2}{\kappa \gamma} \ , 
\end{equation}
assuming $\Omega_{b}$ real. We will assume that $\varepsilon^2 C_b \ , \  (\kappa/\omega_m)^2 C_{b} \ll 1$, which means that the driving of mode $b$ will have a negligible influence on the resonance frequency $\omega_m$, the decay rate $\gamma$, and the noise acting on the mechanical oscillator. Adiabatic elimination of mode $b$ then gives
\begin{align}
\label{eq:OptLangevinAdb}
\dot{a} & = -\frac{\kappa}{2} a - i g_{0} \, x \, a + e^{i \omega_m t} \Omega + \sqrt{\kappa} \, \xi \\
\dot{c} & = -\left(\frac{\gamma}{2} + i \omega_m \right) c - i g_{0} a^\dagger a + e^{-i \omega_m t} 
\Lambda_\mathrm{eff} + \sqrt{\gamma} \, \eta \ .
\label{eq:MechLangevinAdb} 
\end{align}
The only difference from Equations \eqref{eq:OptLangevin} and \eqref{eq:MechLangevin} is the introduction of the effective parameter $\Lambda \rightarrow \Lambda_\mathrm{eff}$, with
\begin{align}
\label{eq:Lambdaeff}
\Lambda_\mathrm{eff} & \approx - 2 i g_{0,b} \varepsilon  \, \chi_{1} \chi_0 \Omega_{b}^2   \ .
\end{align}
We note this is not real as was assumed in the main text (although it is approximately real in the resolved sideband limit), but this has no significance other than a rotation of the quadrature axes.

We can now use our previous results, since $\Lambda_\mathrm{eff}$ plays the role of mechanical drive amplitude. Specifically, we can use Equation \eqref{eq:XEqGen} to determine the mechanical coherent amplitude and phase, as long as we define
\begin{equation}
\label{eq:LDefOpt}
L = \frac{4 g_0 |\Lambda_\mathrm{eff}|}{\gamma \omega_m}  \approx  \frac{g_0}{g_{0,b}} \left(\frac{\kappa}{\omega_m}\right)^2 \varepsilon C_b  \ .
\end{equation}
To realize strong drives $L$ close to the turning point $L_+ \sim C$, we need
\begin{equation}
\label{eq:LDefOpt}
\frac{g_0}{g_{0,b}}  \left(\frac{\kappa}{\omega_m}\right)^2 \varepsilon C_b \sim C  \ .
\end{equation}
With the above assumptions, this roughly translates to
\begin{equation}
\label{eq:LDefOpt}
\frac{g_0}{g_{0,b}} \gg C  \ .
\end{equation}
This shows that to realize the effects we described with optical implementation of the mechanical drive $\Lambda$, the auxiliary cavity mode $b$ must couple significantly weaker to the mechanical oscillator than cavity mode $a$.


\begin{thebibliography}{43}%
\makeatletter
\providecommand \@ifxundefined [1]{%
 \@ifx{#1\undefined}
}%
\providecommand \@ifnum [1]{%
 \ifnum #1\expandafter \@firstoftwo
 \else \expandafter \@secondoftwo
 \fi
}%
\providecommand \@ifx [1]{%
 \ifx #1\expandafter \@firstoftwo
 \else \expandafter \@secondoftwo
 \fi
}%
\providecommand \natexlab [1]{#1}%
\providecommand \enquote  [1]{``#1''}%
\providecommand \bibnamefont  [1]{#1}%
\providecommand \bibfnamefont [1]{#1}%
\providecommand \citenamefont [1]{#1}%
\providecommand \href@noop [0]{\@secondoftwo}%
\providecommand \href [0]{\begingroup \@sanitize@url \@href}%
\providecommand \@href[1]{\@@startlink{#1}\@@href}%
\providecommand \@@href[1]{\endgroup#1\@@endlink}%
\providecommand \@sanitize@url [0]{\catcode `\\12\catcode `\$12\catcode
  `\&12\catcode `\#12\catcode `\^12\catcode `\_12\catcode `\%12\relax}%
\providecommand \@@startlink[1]{}%
\providecommand \@@endlink[0]{}%
\providecommand \url  [0]{\begingroup\@sanitize@url \@url }%
\providecommand \@url [1]{\endgroup\@href {#1}{\urlprefix }}%
\providecommand \urlprefix  [0]{URL }%
\providecommand \Eprint [0]{\href }%
\providecommand \doibase [0]{http://dx.doi.org/}%
\providecommand \selectlanguage [0]{\@gobble}%
\providecommand \bibinfo  [0]{\@secondoftwo}%
\providecommand \bibfield  [0]{\@secondoftwo}%
\providecommand \translation [1]{[#1]}%
\providecommand \BibitemOpen [0]{}%
\providecommand \bibitemStop [0]{}%
\providecommand \bibitemNoStop [0]{.\EOS\space}%
\providecommand \EOS [0]{\spacefactor3000\relax}%
\providecommand \BibitemShut  [1]{\csname bibitem#1\endcsname}%
\let\auto@bib@innerbib\@empty
\bibitem [{\citenamefont {McClelland}\ \emph {et~al.}(2011)\citenamefont
  {McClelland}, \citenamefont {Mavalvala}, \citenamefont {Chen},\ and\
  \citenamefont {Schnabel}}]{McClelland2011LasPhotRev}%
  \BibitemOpen
  \bibfield  {author} {\bibinfo {author} {\bibfnamefont {D.}~\bibnamefont
  {McClelland}}, \bibinfo {author} {\bibfnamefont {N.}~\bibnamefont
  {Mavalvala}}, \bibinfo {author} {\bibfnamefont {Y.}~\bibnamefont {Chen}}, \
  and\ \bibinfo {author} {\bibfnamefont {R.}~\bibnamefont {Schnabel}},\ }\href
  {\doibase 10.1002/lpor.201000034} {\bibfield  {journal} {\bibinfo  {journal}
  {Laser \& Photonics Reviews}\ }\textbf {\bibinfo {volume} {5}},\ \bibinfo
  {pages} {677} (\bibinfo {year} {2011})}\BibitemShut {NoStop}%
\bibitem [{\citenamefont {Metcalfe}(2014)}]{Metcalfe2014ApplPhysRev}%
  \BibitemOpen
  \bibfield  {author} {\bibinfo {author} {\bibfnamefont {M.}~\bibnamefont
  {Metcalfe}},\ }\href {\doibase http://dx.doi.org/10.1063/1.4896029}
  {\bibfield  {journal} {\bibinfo  {journal} {Applied Physics Reviews}\
  }\textbf {\bibinfo {volume} {1}},\ \bibinfo {eid} {031105} (\bibinfo {year}
  {2014})}\BibitemShut {NoStop}%
\bibitem [{\citenamefont {Reed}\ \emph {et~al.}(2017)\citenamefont {Reed},
  \citenamefont {Mayer}, \citenamefont {Teufel}, \citenamefont {Burkhart},
  \citenamefont {Pfaff}, \citenamefont {Reagor}, \citenamefont {Sletten},
  \citenamefont {Ma}, \citenamefont {Schoelkopf}, \citenamefont {Knill},\ and\
  \citenamefont {Lehnert}}]{Reed2017NatPhys}%
  \BibitemOpen
  \bibfield  {author} {\bibinfo {author} {\bibfnamefont {A.~P.}\ \bibnamefont
  {Reed}}, \bibinfo {author} {\bibfnamefont {K.~H.}\ \bibnamefont {Mayer}},
  \bibinfo {author} {\bibfnamefont {J.~D.}\ \bibnamefont {Teufel}}, \bibinfo
  {author} {\bibfnamefont {L.~D.}\ \bibnamefont {Burkhart}}, \bibinfo {author}
  {\bibfnamefont {W.}~\bibnamefont {Pfaff}}, \bibinfo {author} {\bibfnamefont
  {M.}~\bibnamefont {Reagor}}, \bibinfo {author} {\bibfnamefont
  {L.}~\bibnamefont {Sletten}}, \bibinfo {author} {\bibfnamefont
  {X.}~\bibnamefont {Ma}}, \bibinfo {author} {\bibfnamefont {R.~J.}\
  \bibnamefont {Schoelkopf}}, \bibinfo {author} {\bibfnamefont
  {E.}~\bibnamefont {Knill}}, \ and\ \bibinfo {author} {\bibfnamefont {K.~W.}\
  \bibnamefont {Lehnert}},\ }\href@noop {} {\bibfield  {journal} {\bibinfo
  {journal} {Nat. Phys.}\ }\textbf {\bibinfo {volume} {13}},\ \bibinfo {pages}
  {1163} (\bibinfo {year} {2017})}\BibitemShut {NoStop}%
\bibitem [{\citenamefont {Arndt}\ and\ \citenamefont
  {Hornberger}(2014)}]{Arndt2014NatPhys}%
  \BibitemOpen
  \bibfield  {author} {\bibinfo {author} {\bibfnamefont {M.}~\bibnamefont
  {Arndt}}\ and\ \bibinfo {author} {\bibfnamefont {K.}~\bibnamefont
  {Hornberger}},\ }\href@noop {} {\bibfield  {journal} {\bibinfo  {journal}
  {Nat. Phys}\ }\textbf {\bibinfo {volume} {10}},\ \bibinfo {pages} {271}
  (\bibinfo {year} {2014})}\BibitemShut {NoStop}%
\bibitem [{\citenamefont {Teufel}\ \emph {et~al.}(2011)\citenamefont {Teufel},
  \citenamefont {Donner}, \citenamefont {Li}, \citenamefont {Harlow},
  \citenamefont {Allman}, \citenamefont {Cicak}, \citenamefont {Sirois},
  \citenamefont {Whittaker}, \citenamefont {Lehnert},\ and\ \citenamefont
  {Simmonds}}]{Teufel2011Nature}%
  \BibitemOpen
  \bibfield  {author} {\bibinfo {author} {\bibfnamefont {J.~D.}\ \bibnamefont
  {Teufel}}, \bibinfo {author} {\bibfnamefont {T.}~\bibnamefont {Donner}},
  \bibinfo {author} {\bibfnamefont {D.}~\bibnamefont {Li}}, \bibinfo {author}
  {\bibfnamefont {J.~W.}\ \bibnamefont {Harlow}}, \bibinfo {author}
  {\bibfnamefont {M.~S.}\ \bibnamefont {Allman}}, \bibinfo {author}
  {\bibfnamefont {K.}~\bibnamefont {Cicak}}, \bibinfo {author} {\bibfnamefont
  {A.~J.}\ \bibnamefont {Sirois}}, \bibinfo {author} {\bibfnamefont {J.~D.}\
  \bibnamefont {Whittaker}}, \bibinfo {author} {\bibfnamefont {K.~W.}\
  \bibnamefont {Lehnert}}, \ and\ \bibinfo {author} {\bibfnamefont {R.~W.}\
  \bibnamefont {Simmonds}},\ }\href@noop {} {\bibfield  {journal} {\bibinfo
  {journal} {Nature}\ }\textbf {\bibinfo {volume} {475}},\ \bibinfo {pages}
  {359} (\bibinfo {year} {2011})}\BibitemShut {NoStop}%
\bibitem [{\citenamefont {Chan}\ \emph {et~al.}(2011)\citenamefont {Chan},
  \citenamefont {Alegre}, \citenamefont {Safavi-Naeini}, \citenamefont {Hill},
  \citenamefont {Krause}, \citenamefont {Gr\"{o}blacher}, \citenamefont
  {Aspelmeyer},\ and\ \citenamefont {Painter}}]{Chan2011Nature}%
  \BibitemOpen
  \bibfield  {author} {\bibinfo {author} {\bibfnamefont {J.}~\bibnamefont
  {Chan}}, \bibinfo {author} {\bibfnamefont {T.~P.~M.}\ \bibnamefont {Alegre}},
  \bibinfo {author} {\bibfnamefont {A.~H.}\ \bibnamefont {Safavi-Naeini}},
  \bibinfo {author} {\bibfnamefont {J.~T.}\ \bibnamefont {Hill}}, \bibinfo
  {author} {\bibfnamefont {A.}~\bibnamefont {Krause}}, \bibinfo {author}
  {\bibfnamefont {S.}~\bibnamefont {Gr\"{o}blacher}}, \bibinfo {author}
  {\bibfnamefont {M.}~\bibnamefont {Aspelmeyer}}, \ and\ \bibinfo {author}
  {\bibfnamefont {O.}~\bibnamefont {Painter}},\ }\href@noop {} {\bibfield
  {journal} {\bibinfo  {journal} {Nature}\ }\textbf {\bibinfo {volume} {478}},\
  \bibinfo {pages} {89} (\bibinfo {year} {2011})}\BibitemShut {NoStop}%
\bibitem [{\citenamefont {Riedinger}\ \emph {et~al.}(2018)\citenamefont
  {Riedinger}, \citenamefont {Wallucks}, \citenamefont {Marinkovic},
  \citenamefont {L\"{o}schnauer}, \citenamefont {Aspelmeyer}, \citenamefont
  {Hong},\ and\ \citenamefont {Gr\"{o}blacher}}]{Riedinger2018Nature}%
  \BibitemOpen
  \bibfield  {author} {\bibinfo {author} {\bibfnamefont {R.}~\bibnamefont
  {Riedinger}}, \bibinfo {author} {\bibfnamefont {A.}~\bibnamefont {Wallucks}},
  \bibinfo {author} {\bibfnamefont {I.}~\bibnamefont {Marinkovic}}, \bibinfo
  {author} {\bibfnamefont {C.}~\bibnamefont {L\"{o}schnauer}}, \bibinfo
  {author} {\bibfnamefont {M.}~\bibnamefont {Aspelmeyer}}, \bibinfo {author}
  {\bibfnamefont {S.}~\bibnamefont {Hong}}, \ and\ \bibinfo {author}
  {\bibfnamefont {S.}~\bibnamefont {Gr\"{o}blacher}},\ }\href@noop {}
  {\bibfield  {journal} {\bibinfo  {journal} {Nature}\ }\textbf {\bibinfo
  {volume} {556}},\ \bibinfo {pages} {473} (\bibinfo {year}
  {2018})}\BibitemShut {NoStop}%
\bibitem [{\citenamefont {Ockeloen-Korppi}\ \emph {et~al.}(2018)\citenamefont
  {Ockeloen-Korppi}, \citenamefont {Damskagg}, \citenamefont {Pirkkalainen},
  \citenamefont {Clerk}, \citenamefont {Massel}, \citenamefont {Woolley},\ and\
  \citenamefont {Sillanpaa}}]{Ockeloen-Korppi2018Nature}%
  \BibitemOpen
  \bibfield  {author} {\bibinfo {author} {\bibfnamefont {C.~F.}\ \bibnamefont
  {Ockeloen-Korppi}}, \bibinfo {author} {\bibfnamefont {E.}~\bibnamefont
  {Damskagg}}, \bibinfo {author} {\bibfnamefont {J.-M.}\ \bibnamefont
  {Pirkkalainen}}, \bibinfo {author} {\bibfnamefont {A.~A.}\ \bibnamefont
  {Clerk}}, \bibinfo {author} {\bibfnamefont {F.}~\bibnamefont {Massel}},
  \bibinfo {author} {\bibfnamefont {M.~J.}\ \bibnamefont {Woolley}}, \ and\
  \bibinfo {author} {\bibfnamefont {M.~A.}\ \bibnamefont {Sillanpaa}},\
  }\href@noop {} {\bibfield  {journal} {\bibinfo  {journal} {Nature}\ }\textbf
  {\bibinfo {volume} {556}},\ \bibinfo {pages} {478} (\bibinfo {year}
  {2018})}\BibitemShut {NoStop}%
\bibitem [{\citenamefont {Verhagen}\ and\ \citenamefont
  {Al\'{u}}(2017)}]{Verhagen2017NatPhys}%
  \BibitemOpen
  \bibfield  {author} {\bibinfo {author} {\bibfnamefont {E.}~\bibnamefont
  {Verhagen}}\ and\ \bibinfo {author} {\bibfnamefont {A.}~\bibnamefont
  {Al\'{u}}},\ }\href@noop {} {\bibfield  {journal} {\bibinfo  {journal} {Nat.
  Phys.}\ }\textbf {\bibinfo {volume} {13}},\ \bibinfo {pages} {922} (\bibinfo
  {year} {2017})}\BibitemShut {NoStop}%
\bibitem [{\citenamefont {Braginsky}\ \emph {et~al.}(2001)\citenamefont
  {Braginsky}, \citenamefont {Strigin},\ and\ \citenamefont
  {Vyatchanin}}]{Braginsky2001PhysLettA}%
  \BibitemOpen
  \bibfield  {author} {\bibinfo {author} {\bibfnamefont {V.~B.}\ \bibnamefont
  {Braginsky}}, \bibinfo {author} {\bibfnamefont {S.~E.}\ \bibnamefont
  {Strigin}}, \ and\ \bibinfo {author} {\bibfnamefont {S.~P.}\ \bibnamefont
  {Vyatchanin}},\ }\href@noop {} {\bibfield  {journal} {\bibinfo  {journal}
  {Phys. Lett. A}\ }\textbf {\bibinfo {volume} {287}},\ \bibinfo {pages} {331}
  (\bibinfo {year} {2001})}\BibitemShut {NoStop}%
\bibitem [{\citenamefont {Carmon}\ \emph {et~al.}(2005)\citenamefont {Carmon},
  \citenamefont {Rokhsari}, \citenamefont {Yang}, \citenamefont {Kippenberg},\
  and\ \citenamefont {Vahala}}]{Carmon2005PRL}%
  \BibitemOpen
  \bibfield  {author} {\bibinfo {author} {\bibfnamefont {T.}~\bibnamefont
  {Carmon}}, \bibinfo {author} {\bibfnamefont {H.}~\bibnamefont {Rokhsari}},
  \bibinfo {author} {\bibfnamefont {L.}~\bibnamefont {Yang}}, \bibinfo {author}
  {\bibfnamefont {T.~J.}\ \bibnamefont {Kippenberg}}, \ and\ \bibinfo {author}
  {\bibfnamefont {K.~J.}\ \bibnamefont {Vahala}},\ }\href {\doibase
  10.1103/PhysRevLett.94.223902} {\bibfield  {journal} {\bibinfo  {journal}
  {Phys. Rev. Lett.}\ }\textbf {\bibinfo {volume} {94}},\ \bibinfo {pages}
  {223902} (\bibinfo {year} {2005})}\BibitemShut {NoStop}%
\bibitem [{\citenamefont {Kippenberg}\ \emph {et~al.}(2005)\citenamefont
  {Kippenberg}, \citenamefont {Rokhsari}, \citenamefont {Carmon}, \citenamefont
  {Scherer},\ and\ \citenamefont {Vahala}}]{Kippenberg2005PRL}%
  \BibitemOpen
  \bibfield  {author} {\bibinfo {author} {\bibfnamefont {T.~J.}\ \bibnamefont
  {Kippenberg}}, \bibinfo {author} {\bibfnamefont {H.}~\bibnamefont
  {Rokhsari}}, \bibinfo {author} {\bibfnamefont {T.}~\bibnamefont {Carmon}},
  \bibinfo {author} {\bibfnamefont {A.}~\bibnamefont {Scherer}}, \ and\
  \bibinfo {author} {\bibfnamefont {K.~J.}\ \bibnamefont {Vahala}},\ }\href
  {\doibase 10.1103/PhysRevLett.95.033901} {\bibfield  {journal} {\bibinfo
  {journal} {Phys. Rev. Lett.}\ }\textbf {\bibinfo {volume} {95}},\ \bibinfo
  {pages} {033901} (\bibinfo {year} {2005})}\BibitemShut {NoStop}%
\bibitem [{\citenamefont {Marquardt}\ \emph {et~al.}(2006)\citenamefont
  {Marquardt}, \citenamefont {Harris},\ and\ \citenamefont
  {Girvin}}]{Marquardt2006PRL}%
  \BibitemOpen
  \bibfield  {author} {\bibinfo {author} {\bibfnamefont {F.}~\bibnamefont
  {Marquardt}}, \bibinfo {author} {\bibfnamefont {J.~G.~E.}\ \bibnamefont
  {Harris}}, \ and\ \bibinfo {author} {\bibfnamefont {S.~M.}\ \bibnamefont
  {Girvin}},\ }\href {\doibase 10.1103/PhysRevLett.96.103901} {\bibfield
  {journal} {\bibinfo  {journal} {Phys. Rev. Lett.}\ }\textbf {\bibinfo
  {volume} {96}},\ \bibinfo {pages} {103901} (\bibinfo {year}
  {2006})}\BibitemShut {NoStop}%
\bibitem [{\citenamefont {Krause}\ \emph {et~al.}(2015)\citenamefont {Krause},
  \citenamefont {Hill}, \citenamefont {Ludwig}, \citenamefont {Safavi-Naeini},
  \citenamefont {Chan}, \citenamefont {Marquardt},\ and\ \citenamefont
  {Painter}}]{Krause2015PRL}%
  \BibitemOpen
  \bibfield  {author} {\bibinfo {author} {\bibfnamefont {A.~G.}\ \bibnamefont
  {Krause}}, \bibinfo {author} {\bibfnamefont {J.~T.}\ \bibnamefont {Hill}},
  \bibinfo {author} {\bibfnamefont {M.}~\bibnamefont {Ludwig}}, \bibinfo
  {author} {\bibfnamefont {A.~H.}\ \bibnamefont {Safavi-Naeini}}, \bibinfo
  {author} {\bibfnamefont {J.}~\bibnamefont {Chan}}, \bibinfo {author}
  {\bibfnamefont {F.}~\bibnamefont {Marquardt}}, \ and\ \bibinfo {author}
  {\bibfnamefont {O.}~\bibnamefont {Painter}},\ }\href {\doibase
  10.1103/PhysRevLett.115.233601} {\bibfield  {journal} {\bibinfo  {journal}
  {Phys. Rev. Lett.}\ }\textbf {\bibinfo {volume} {115}},\ \bibinfo {pages}
  {233601} (\bibinfo {year} {2015})}\BibitemShut {NoStop}%
\bibitem [{\citenamefont {Buters}\ \emph {et~al.}(2015)\citenamefont {Buters},
  \citenamefont {Eerkens}, \citenamefont {Heeck}, \citenamefont {Weaver},
  \citenamefont {Pepper}, \citenamefont {de~Man},\ and\ \citenamefont
  {Bouwmeester}}]{Buters2015PRA}%
  \BibitemOpen
  \bibfield  {author} {\bibinfo {author} {\bibfnamefont {F.~M.}\ \bibnamefont
  {Buters}}, \bibinfo {author} {\bibfnamefont {H.~J.}\ \bibnamefont {Eerkens}},
  \bibinfo {author} {\bibfnamefont {K.}~\bibnamefont {Heeck}}, \bibinfo
  {author} {\bibfnamefont {M.~J.}\ \bibnamefont {Weaver}}, \bibinfo {author}
  {\bibfnamefont {B.}~\bibnamefont {Pepper}}, \bibinfo {author} {\bibfnamefont
  {S.}~\bibnamefont {de~Man}}, \ and\ \bibinfo {author} {\bibfnamefont
  {D.}~\bibnamefont {Bouwmeester}},\ }\href {\doibase
  10.1103/PhysRevA.92.013811} {\bibfield  {journal} {\bibinfo  {journal} {Phys.
  Rev. A}\ }\textbf {\bibinfo {volume} {92}},\ \bibinfo {pages} {013811}
  (\bibinfo {year} {2015})}\BibitemShut {NoStop}%
\bibitem [{\citenamefont {Gibbs}\ \emph {et~al.}(1976)\citenamefont {Gibbs},
  \citenamefont {McCall},\ and\ \citenamefont {Venkatesan}}]{Gibbs1976PRL}%
  \BibitemOpen
  \bibfield  {author} {\bibinfo {author} {\bibfnamefont {H.~M.}\ \bibnamefont
  {Gibbs}}, \bibinfo {author} {\bibfnamefont {S.~L.}\ \bibnamefont {McCall}}, \
  and\ \bibinfo {author} {\bibfnamefont {T.~N.~C.}\ \bibnamefont
  {Venkatesan}},\ }\href {\doibase 10.1103/PhysRevLett.36.1135} {\bibfield
  {journal} {\bibinfo  {journal} {Phys. Rev. Lett.}\ }\textbf {\bibinfo
  {volume} {36}},\ \bibinfo {pages} {1135} (\bibinfo {year}
  {1976})}\BibitemShut {NoStop}%
\bibitem [{\citenamefont {Siddiqi}\ \emph {et~al.}(2004)\citenamefont
  {Siddiqi}, \citenamefont {Vijay}, \citenamefont {Pierre}, \citenamefont
  {Wilson}, \citenamefont {Metcalfe}, \citenamefont {Rigetti}, \citenamefont
  {Frunzio},\ and\ \citenamefont {Devoret}}]{Siddiqi2004PRL}%
  \BibitemOpen
  \bibfield  {author} {\bibinfo {author} {\bibfnamefont {I.}~\bibnamefont
  {Siddiqi}}, \bibinfo {author} {\bibfnamefont {R.}~\bibnamefont {Vijay}},
  \bibinfo {author} {\bibfnamefont {F.}~\bibnamefont {Pierre}}, \bibinfo
  {author} {\bibfnamefont {C.~M.}\ \bibnamefont {Wilson}}, \bibinfo {author}
  {\bibfnamefont {M.}~\bibnamefont {Metcalfe}}, \bibinfo {author}
  {\bibfnamefont {C.}~\bibnamefont {Rigetti}}, \bibinfo {author} {\bibfnamefont
  {L.}~\bibnamefont {Frunzio}}, \ and\ \bibinfo {author} {\bibfnamefont
  {M.~H.}\ \bibnamefont {Devoret}},\ }\href {\doibase
  10.1103/PhysRevLett.93.207002} {\bibfield  {journal} {\bibinfo  {journal}
  {Phys. Rev. Lett.}\ }\textbf {\bibinfo {volume} {93}},\ \bibinfo {pages}
  {207002} (\bibinfo {year} {2004})}\BibitemShut {NoStop}%
\bibitem [{\citenamefont {Siddiqi}\ \emph {et~al.}(2005)\citenamefont
  {Siddiqi}, \citenamefont {Vijay}, \citenamefont {Pierre}, \citenamefont
  {Wilson}, \citenamefont {Frunzio}, \citenamefont {Metcalfe}, \citenamefont
  {Rigetti}, \citenamefont {Schoelkopf}, \citenamefont {Devoret}, \citenamefont
  {Vion},\ and\ \citenamefont {Esteve}}]{Siddiqi2005PRL}%
  \BibitemOpen
  \bibfield  {author} {\bibinfo {author} {\bibfnamefont {I.}~\bibnamefont
  {Siddiqi}}, \bibinfo {author} {\bibfnamefont {R.}~\bibnamefont {Vijay}},
  \bibinfo {author} {\bibfnamefont {F.}~\bibnamefont {Pierre}}, \bibinfo
  {author} {\bibfnamefont {C.~M.}\ \bibnamefont {Wilson}}, \bibinfo {author}
  {\bibfnamefont {L.}~\bibnamefont {Frunzio}}, \bibinfo {author} {\bibfnamefont
  {M.}~\bibnamefont {Metcalfe}}, \bibinfo {author} {\bibfnamefont
  {C.}~\bibnamefont {Rigetti}}, \bibinfo {author} {\bibfnamefont {R.~J.}\
  \bibnamefont {Schoelkopf}}, \bibinfo {author} {\bibfnamefont {M.~H.}\
  \bibnamefont {Devoret}}, \bibinfo {author} {\bibfnamefont {D.}~\bibnamefont
  {Vion}}, \ and\ \bibinfo {author} {\bibfnamefont {D.}~\bibnamefont
  {Esteve}},\ }\href {\doibase 10.1103/PhysRevLett.94.027005} {\bibfield
  {journal} {\bibinfo  {journal} {Phys. Rev. Lett.}\ }\textbf {\bibinfo
  {volume} {94}},\ \bibinfo {pages} {027005} (\bibinfo {year}
  {2005})}\BibitemShut {NoStop}%
\bibitem [{\citenamefont {Dykman}\ and\ \citenamefont
  {Krivoglaz}(1980)}]{Dykman1980SovPhysJETP}%
  \BibitemOpen
  \bibfield  {author} {\bibinfo {author} {\bibfnamefont {M.~I.}\ \bibnamefont
  {Dykman}}\ and\ \bibinfo {author} {\bibfnamefont {M.~A.}\ \bibnamefont
  {Krivoglaz}},\ }\href@noop {} {\bibfield  {journal} {\bibinfo  {journal} {Zh.
  Eksp. Teor. Fi.}\ }\textbf {\bibinfo {volume} {77}},\ \bibinfo {pages} {60}
  (\bibinfo {year} {1980})},\ \bibinfo {note} {[Sov. Phys. JETP {\bf 50}, 30
  (1979)]}\BibitemShut {NoStop}%
\bibitem [{\citenamefont {Aldridge}\ and\ \citenamefont
  {Cleland}(2005)}]{Aldridge2005PRL}%
  \BibitemOpen
  \bibfield  {author} {\bibinfo {author} {\bibfnamefont {J.~S.}\ \bibnamefont
  {Aldridge}}\ and\ \bibinfo {author} {\bibfnamefont {A.~N.}\ \bibnamefont
  {Cleland}},\ }\href {\doibase 10.1103/PhysRevLett.94.156403} {\bibfield
  {journal} {\bibinfo  {journal} {Phys. Rev. Lett.}\ }\textbf {\bibinfo
  {volume} {94}},\ \bibinfo {pages} {156403} (\bibinfo {year}
  {2005})}\BibitemShut {NoStop}%
\bibitem [{\citenamefont {Karabalin}\ \emph {et~al.}(2011)\citenamefont
  {Karabalin}, \citenamefont {Lifshitz}, \citenamefont {Cross}, \citenamefont
  {Matheny}, \citenamefont {Masmanidis},\ and\ \citenamefont
  {Roukes}}]{Karabalin2011PRL}%
  \BibitemOpen
  \bibfield  {author} {\bibinfo {author} {\bibfnamefont {R.~B.}\ \bibnamefont
  {Karabalin}}, \bibinfo {author} {\bibfnamefont {R.}~\bibnamefont {Lifshitz}},
  \bibinfo {author} {\bibfnamefont {M.~C.}\ \bibnamefont {Cross}}, \bibinfo
  {author} {\bibfnamefont {M.~H.}\ \bibnamefont {Matheny}}, \bibinfo {author}
  {\bibfnamefont {S.~C.}\ \bibnamefont {Masmanidis}}, \ and\ \bibinfo {author}
  {\bibfnamefont {M.~L.}\ \bibnamefont {Roukes}},\ }\href {\doibase
  10.1103/PhysRevLett.106.094102} {\bibfield  {journal} {\bibinfo  {journal}
  {Phys. Rev. Lett.}\ }\textbf {\bibinfo {volume} {106}},\ \bibinfo {pages}
  {094102} (\bibinfo {year} {2011})}\BibitemShut {NoStop}%
\bibitem [{\citenamefont {Dong}\ \emph {et~al.}(2018)\citenamefont {Dong},
  \citenamefont {Dykman},\ and\ \citenamefont {Chan}}]{Dong2018NatComm}%
  \BibitemOpen
  \bibfield  {author} {\bibinfo {author} {\bibfnamefont {X.}~\bibnamefont
  {Dong}}, \bibinfo {author} {\bibfnamefont {M.}~\bibnamefont {Dykman}}, \ and\
  \bibinfo {author} {\bibfnamefont {H.}~\bibnamefont {Chan}},\ }\href@noop {}
  {\bibfield  {journal} {\bibinfo  {journal} {Nat. Comm.}\ }\textbf {\bibinfo
  {volume} {9}},\ \bibinfo {pages}
  {3241} (\bibinfo {year} {2018})}\BibitemShut {NoStop}%
\bibitem [{\citenamefont {Dorsel}\ \emph {et~al.}(1983)\citenamefont {Dorsel},
  \citenamefont {McCullen}, \citenamefont {Meystre}, \citenamefont {Vignes},\
  and\ \citenamefont {Walther}}]{Dorsel1983PRL}%
  \BibitemOpen
  \bibfield  {author} {\bibinfo {author} {\bibfnamefont {A.}~\bibnamefont
  {Dorsel}}, \bibinfo {author} {\bibfnamefont {J.~D.}\ \bibnamefont
  {McCullen}}, \bibinfo {author} {\bibfnamefont {P.}~\bibnamefont {Meystre}},
  \bibinfo {author} {\bibfnamefont {E.}~\bibnamefont {Vignes}}, \ and\ \bibinfo
  {author} {\bibfnamefont {H.}~\bibnamefont {Walther}},\ }\href {\doibase
  10.1103/PhysRevLett.51.1550} {\bibfield  {journal} {\bibinfo  {journal}
  {Phys. Rev. Lett.}\ }\textbf {\bibinfo {volume} {51}},\ \bibinfo {pages}
  {1550} (\bibinfo {year} {1983})}\BibitemShut {NoStop}%
\bibitem [{\citenamefont {Bagheri}\ \emph {et~al.}(2011)\citenamefont
  {Bagheri}, \citenamefont {Poot}, \citenamefont {Li}, \citenamefont
  {Pernice},\ and\ \citenamefont {Tang}}]{Bagheri2011NatNano}%
  \BibitemOpen
  \bibfield  {author} {\bibinfo {author} {\bibfnamefont {M.}~\bibnamefont
  {Bagheri}}, \bibinfo {author} {\bibfnamefont {M.}~\bibnamefont {Poot}},
  \bibinfo {author} {\bibfnamefont {M.}~\bibnamefont {Li}}, \bibinfo {author}
  {\bibfnamefont {W.~P.~H.}\ \bibnamefont {Pernice}}, \ and\ \bibinfo {author}
  {\bibfnamefont {H.~X.}\ \bibnamefont {Tang}},\ }\href@noop {} {\bibfield
  {journal} {\bibinfo  {journal} {Nature Nanotech.}\ }\textbf {\bibinfo
  {volume} {6}},\ \bibinfo {pages} {726} (\bibinfo {year} {2011})}\BibitemShut
  {NoStop}%
\bibitem [{\citenamefont {Xu}\ \emph {et~al.}(2017)\citenamefont {Xu},
  \citenamefont {Kemiktarak}, \citenamefont {Fan}, \citenamefont {Ragole},
  \citenamefont {Lawall},\ and\ \citenamefont {Taylor}}]{Xu2017NatComm}%
  \BibitemOpen
  \bibfield  {author} {\bibinfo {author} {\bibfnamefont {H.}~\bibnamefont
  {Xu}}, \bibinfo {author} {\bibfnamefont {U.}~\bibnamefont {Kemiktarak}},
  \bibinfo {author} {\bibfnamefont {J.}~\bibnamefont {Fan}}, \bibinfo {author}
  {\bibfnamefont {S.}~\bibnamefont {Ragole}}, \bibinfo {author} {\bibfnamefont
  {J.}~\bibnamefont {Lawall}}, \ and\ \bibinfo {author} {\bibfnamefont {J.~M.}\
  \bibnamefont {Taylor}},\ }\href@noop {} {\bibfield  {journal} {\bibinfo
  {journal} {Nat. Comm.}\ }\textbf {\bibinfo {volume} {8}},\ \bibinfo {pages}
  {14481} (\bibinfo {year}
  {2017})}\BibitemShut {NoStop}%
\bibitem [{\citenamefont {Dykman}(2007)}]{Dykman2007PRE}%
  \BibitemOpen
  \bibfield  {author} {\bibinfo {author} {\bibfnamefont {M.~I.}\ \bibnamefont
  {Dykman}},\ }\href {\doibase 10.1103/PhysRevE.75.011101} {\bibfield
  {journal} {\bibinfo  {journal} {Phys. Rev. E}\ }\textbf {\bibinfo {volume}
  {75}},\ \bibinfo {pages} {011101} (\bibinfo {year} {2007})}\BibitemShut
  {NoStop}%
\bibitem [{\citenamefont {Gardiner}\ and\ \citenamefont
  {Collett}(1985)}]{Gardiner1985PRA}%
  \BibitemOpen
  \bibfield  {author} {\bibinfo {author} {\bibfnamefont {C.~W.}\ \bibnamefont
  {Gardiner}}\ and\ \bibinfo {author} {\bibfnamefont {M.~J.}\ \bibnamefont
  {Collett}},\ }\href {\doibase 10.1103/PhysRevA.31.3761} {\bibfield  {journal}
  {\bibinfo  {journal} {Phys. Rev. A}\ }\textbf {\bibinfo {volume} {31}},\
  \bibinfo {pages} {3761} (\bibinfo {year} {1985})}\BibitemShut {NoStop}%
\bibitem [{\citenamefont {Clerk}\ \emph {et~al.}(2010)\citenamefont {Clerk},
  \citenamefont {Devoret}, \citenamefont {Girvin}, \citenamefont {Marquardt},\
  and\ \citenamefont {Schoelkopf}}]{Clerk2010RMP}%
  \BibitemOpen
  \bibfield  {author} {\bibinfo {author} {\bibfnamefont {A.~A.}\ \bibnamefont
  {Clerk}}, \bibinfo {author} {\bibfnamefont {M.~H.}\ \bibnamefont {Devoret}},
  \bibinfo {author} {\bibfnamefont {S.~M.}\ \bibnamefont {Girvin}}, \bibinfo
  {author} {\bibfnamefont {F.}~\bibnamefont {Marquardt}}, \ and\ \bibinfo
  {author} {\bibfnamefont {R.~J.}\ \bibnamefont {Schoelkopf}},\ }\href
  {\doibase 10.1103/RevModPhys.82.1155} {\bibfield  {journal} {\bibinfo
  {journal} {Rev. Mod. Phys.}\ }\textbf {\bibinfo {volume} {82}},\ \bibinfo
  {pages} {1155} (\bibinfo {year} {2010})}\BibitemShut {NoStop}%
\bibitem [{\citenamefont {Braginsky}\ and\ \citenamefont
  {Vyatchanin}(2002)}]{Braginsky2002PhysLettA}%
  \BibitemOpen
  \bibfield  {author} {\bibinfo {author} {\bibfnamefont {V.~B.}\ \bibnamefont
  {Braginsky}}\ and\ \bibinfo {author} {\bibfnamefont {S.~P.}\ \bibnamefont
  {Vyatchanin}},\ }\href@noop {} {\bibfield  {journal} {\bibinfo  {journal}
  {Phys. Lett. A}\ }\textbf {\bibinfo {volume} {293}},\ \bibinfo {pages} {228}
  (\bibinfo {year} {2002})}\BibitemShut {NoStop}%
\bibitem [{\citenamefont {Marquardt}\ \emph {et~al.}(2007)\citenamefont
  {Marquardt}, \citenamefont {Chen}, \citenamefont {Clerk},\ and\ \citenamefont
  {Girvin}}]{Marquardt2007PRL}%
  \BibitemOpen
  \bibfield  {author} {\bibinfo {author} {\bibfnamefont {F.}~\bibnamefont
  {Marquardt}}, \bibinfo {author} {\bibfnamefont {J.~P.}\ \bibnamefont {Chen}},
  \bibinfo {author} {\bibfnamefont {A.~A.}\ \bibnamefont {Clerk}}, \ and\
  \bibinfo {author} {\bibfnamefont {S.~M.}\ \bibnamefont {Girvin}},\ }\href
  {\doibase 10.1103/PhysRevLett.99.093902} {\bibfield  {journal} {\bibinfo
  {journal} {Phys. Rev. Lett.}\ }\textbf {\bibinfo {volume} {99}},\ \bibinfo
  {pages} {093902} (\bibinfo {year} {2007})}\BibitemShut {NoStop}%
\bibitem [{\citenamefont {Wilson-Rae}\ \emph {et~al.}(2007)\citenamefont
  {Wilson-Rae}, \citenamefont {Nooshi}, \citenamefont {Zwerger},\ and\
  \citenamefont {Kippenberg}}]{Wilson-Rae2007PRL}%
  \BibitemOpen
  \bibfield  {author} {\bibinfo {author} {\bibfnamefont {I.}~\bibnamefont
  {Wilson-Rae}}, \bibinfo {author} {\bibfnamefont {N.}~\bibnamefont {Nooshi}},
  \bibinfo {author} {\bibfnamefont {W.}~\bibnamefont {Zwerger}}, \ and\
  \bibinfo {author} {\bibfnamefont {T.~J.}\ \bibnamefont {Kippenberg}},\ }\href
  {\doibase 10.1103/PhysRevLett.99.093901} {\bibfield  {journal} {\bibinfo
  {journal} {Phys. Rev. Lett.}\ }\textbf {\bibinfo {volume} {99}},\ \bibinfo
  {pages} {093901} (\bibinfo {year} {2007})}\BibitemShut {NoStop}%
\bibitem [{\citenamefont {Kronwald}\ \emph {et~al.}(2013)\citenamefont
  {Kronwald}, \citenamefont {Marquardt},\ and\ \citenamefont
  {Clerk}}]{Kronwald2013PRA_2}%
  \BibitemOpen
  \bibfield  {author} {\bibinfo {author} {\bibfnamefont {A.}~\bibnamefont
  {Kronwald}}, \bibinfo {author} {\bibfnamefont {F.}~\bibnamefont {Marquardt}},
  \ and\ \bibinfo {author} {\bibfnamefont {A.~A.}\ \bibnamefont {Clerk}},\
  }\href {\doibase 10.1103/PhysRevA.88.063833} {\bibfield  {journal} {\bibinfo
  {journal} {Phys. Rev. A}\ }\textbf {\bibinfo {volume} {88}},\ \bibinfo
  {pages} {063833} (\bibinfo {year} {2013})}\BibitemShut {NoStop}%
\bibitem [{\citenamefont {Kramers}(1940)}]{Kramers1940Physica}%
  \BibitemOpen
  \bibfield  {author} {\bibinfo {author} {\bibfnamefont {H.}~\bibnamefont
  {Kramers}},\ }\href@noop {} {\bibfield  {journal} {\bibinfo  {journal}
  {Physica (Utrecht)}\ }\textbf {\bibinfo {volume} {7}},\ \bibinfo {pages}
  {284} (\bibinfo {year} {1940})}\BibitemShut {NoStop}%
\bibitem [{\citenamefont {Langer}(1969)}]{Langer1969AnnPhys}%
  \BibitemOpen
  \bibfield  {author} {\bibinfo {author} {\bibfnamefont {J.}~\bibnamefont
  {Langer}},\ }\href {\doibase https://doi.org/10.1016/0003-4916(69)90153-5}
  {\bibfield  {journal} {\bibinfo  {journal} {Annals of Physics}\ }\textbf
  {\bibinfo {volume} {54}},\ \bibinfo {pages} {258 } (\bibinfo {year}
  {1969})}\BibitemShut {NoStop}%
\bibitem [{\citenamefont {H\"{a}nggi}(1986)}]{Hanggi1986JStatPhys}%
  \BibitemOpen
  \bibfield  {author} {\bibinfo {author} {\bibfnamefont {P.}~\bibnamefont
  {H\"{a}nggi}},\ }\href@noop {} {\bibfield  {journal} {\bibinfo  {journal}
  {Journal of Statistical Physics}\ }\textbf {\bibinfo {volume} {42}},\
  \bibinfo {pages} {105} (\bibinfo {year} {1986})}\BibitemShut {NoStop}%
\bibitem [{\citenamefont {Kloeden}\ and\ \citenamefont
  {Platen}(1999)}]{Kloeden1999Book}%
  \BibitemOpen
  \bibfield  {author} {\bibinfo {author} {\bibfnamefont {P.~E.}\ \bibnamefont
  {Kloeden}}\ and\ \bibinfo {author} {\bibfnamefont {E.}~\bibnamefont
  {Platen}},\ }\href@noop {} {\emph {\bibinfo {title} {Numerical Solution to
  Stochastic Differential Equations}}}\ (\bibinfo  {publisher} {Springer, New
  York},\ \bibinfo {year} {1999})\BibitemShut {NoStop}%
\bibitem [{\citenamefont {Heinrich}\ \emph {et~al.}(2010)\citenamefont
  {Heinrich}, \citenamefont {Harris},\ and\ \citenamefont
  {Marquardt}}]{Heinrich2010PRA}%
  \BibitemOpen
  \bibfield  {author} {\bibinfo {author} {\bibfnamefont {G.}~\bibnamefont
  {Heinrich}}, \bibinfo {author} {\bibfnamefont {J.~G.~E.}\ \bibnamefont
  {Harris}}, \ and\ \bibinfo {author} {\bibfnamefont {F.}~\bibnamefont
  {Marquardt}},\ }\href {\doibase 10.1103/PhysRevA.81.011801} {\bibfield
  {journal} {\bibinfo  {journal} {Phys. Rev. A}\ }\textbf {\bibinfo {volume}
  {81}},\ \bibinfo {pages} {011801} (\bibinfo {year} {2010})}\BibitemShut
  {NoStop}%
\bibitem [{\citenamefont {Safavi-Naeini}\ and\ \citenamefont
  {Painter}(2011)}]{Safavi-Naeini2011NJP}%
  \BibitemOpen
  \bibfield  {author} {\bibinfo {author} {\bibfnamefont {A.~H.}\ \bibnamefont
  {Safavi-Naeini}}\ and\ \bibinfo {author} {\bibfnamefont {O.}~\bibnamefont
  {Painter}},\ }\href {http://stacks.iop.org/1367-2630/13/i=1/a=013017}
  {\bibfield  {journal} {\bibinfo  {journal} {New Journal of Physics}\ }\textbf
  {\bibinfo {volume} {13}},\ \bibinfo {pages} {013017} (\bibinfo {year}
  {2011})}\BibitemShut {NoStop}%
\bibitem [{\citenamefont {Ludwig}\ \emph {et~al.}(2012)\citenamefont {Ludwig},
  \citenamefont {Safavi-Naeini}, \citenamefont {Painter},\ and\ \citenamefont
  {Marquardt}}]{Ludwig2012PRL}%
  \BibitemOpen
  \bibfield  {author} {\bibinfo {author} {\bibfnamefont {M.}~\bibnamefont
  {Ludwig}}, \bibinfo {author} {\bibfnamefont {A.~H.}\ \bibnamefont
  {Safavi-Naeini}}, \bibinfo {author} {\bibfnamefont {O.}~\bibnamefont
  {Painter}}, \ and\ \bibinfo {author} {\bibfnamefont {F.}~\bibnamefont
  {Marquardt}},\ }\href {\doibase 10.1103/PhysRevLett.109.063601} {\bibfield
  {journal} {\bibinfo  {journal} {Phys. Rev. Lett.}\ }\textbf {\bibinfo
  {volume} {109}},\ \bibinfo {pages} {063601} (\bibinfo {year}
  {2012})}\BibitemShut {NoStop}%
\bibitem [{\citenamefont {Stannigel}\ \emph {et~al.}(2012)\citenamefont
  {Stannigel}, \citenamefont {Komar}, \citenamefont {Habraken}, \citenamefont
  {Bennett}, \citenamefont {Lukin}, \citenamefont {Zoller},\ and\ \citenamefont
  {Rabl}}]{Stannigel2012PRL}%
  \BibitemOpen
  \bibfield  {author} {\bibinfo {author} {\bibfnamefont {K.}~\bibnamefont
  {Stannigel}}, \bibinfo {author} {\bibfnamefont {P.}~\bibnamefont {Komar}},
  \bibinfo {author} {\bibfnamefont {S.~J.~M.}\ \bibnamefont {Habraken}},
  \bibinfo {author} {\bibfnamefont {S.~D.}\ \bibnamefont {Bennett}}, \bibinfo
  {author} {\bibfnamefont {M.~D.}\ \bibnamefont {Lukin}}, \bibinfo {author}
  {\bibfnamefont {P.}~\bibnamefont {Zoller}}, \ and\ \bibinfo {author}
  {\bibfnamefont {P.}~\bibnamefont {Rabl}},\ }\href {\doibase
  10.1103/PhysRevLett.109.013603} {\bibfield  {journal} {\bibinfo  {journal}
  {Phys. Rev. Lett.}\ }\textbf {\bibinfo {volume} {109}},\ \bibinfo {pages}
  {013603} (\bibinfo {year} {2012})}\BibitemShut {NoStop}%
\bibitem [{\citenamefont {Thompson}\ \emph {et~al.}(2008)\citenamefont
  {Thompson}, \citenamefont {Zwickl}, \citenamefont {Jayich}, \citenamefont
  {Marquardt}, \citenamefont {Girvin},\ and\ \citenamefont
  {Harris}}]{Thompson2008Nature}%
  \BibitemOpen
  \bibfield  {author} {\bibinfo {author} {\bibfnamefont {J.~D.}\ \bibnamefont
  {Thompson}}, \bibinfo {author} {\bibfnamefont {B.~M.}\ \bibnamefont
  {Zwickl}}, \bibinfo {author} {\bibfnamefont {A.~M.}\ \bibnamefont {Jayich}},
  \bibinfo {author} {\bibfnamefont {F.}~\bibnamefont {Marquardt}}, \bibinfo
  {author} {\bibfnamefont {S.~M.}\ \bibnamefont {Girvin}}, \ and\ \bibinfo
  {author} {\bibfnamefont {J.~G.~E.}\ \bibnamefont {Harris}},\ }\href {\doibase
  10.1038/nature06715} {\bibfield  {journal} {\bibinfo  {journal} {Nature}\
  }\textbf {\bibinfo {volume} {452}},\ \bibinfo {pages} {72} (\bibinfo {year}
  {2008})}\BibitemShut {NoStop}%
\bibitem [{\citenamefont {Eichenfield}\ \emph {et~al.}(2009)\citenamefont
  {Eichenfield}, \citenamefont {Chan}, \citenamefont {Camacho}, \citenamefont
  {Vahala},\ and\ \citenamefont {Painter}}]{Eichenfield2009Nature_2}%
  \BibitemOpen
  \bibfield  {author} {\bibinfo {author} {\bibfnamefont {M.}~\bibnamefont
  {Eichenfield}}, \bibinfo {author} {\bibfnamefont {J.}~\bibnamefont {Chan}},
  \bibinfo {author} {\bibfnamefont {R.~M.}\ \bibnamefont {Camacho}}, \bibinfo
  {author} {\bibfnamefont {K.~J.}\ \bibnamefont {Vahala}}, \ and\ \bibinfo
  {author} {\bibfnamefont {O.}~\bibnamefont {Painter}},\ }\href@noop {}
  {\bibfield  {journal} {\bibinfo  {journal} {Nature}\ }\textbf {\bibinfo
  {volume} {462}},\ \bibinfo {pages} {78} (\bibinfo {year} {2009})}\BibitemShut
  {NoStop}%
\bibitem [{\citenamefont {Autler}\ and\ \citenamefont
  {Townes}(1955)}]{Autler1955PR}%
  \BibitemOpen
  \bibfield  {author} {\bibinfo {author} {\bibfnamefont {S.~H.}\ \bibnamefont
  {Autler}}\ and\ \bibinfo {author} {\bibfnamefont {C.~H.}\ \bibnamefont
  {Townes}},\ }\href {\doibase 10.1103/PhysRev.100.703} {\bibfield  {journal}
  {\bibinfo  {journal} {Phys. Rev.}\ }\textbf {\bibinfo {volume} {100}},\
  \bibinfo {pages} {703} (\bibinfo {year} {1955})}\BibitemShut {NoStop}%
\end{thebibliography}
%

\end{document}